\documentclass[reprint,nofootinbib,amsmath,amssymb,aps]{revtex4-2}
\pdfoutput=1
\usepackage{graphicx}
\usepackage{dcolumn}
\usepackage{bm}
\usepackage{hyperref}

\usepackage{lipsum}

\newcommand{\dif}{\mathrm{d}}
\newcommand{\const}{\mathrm{const.}}
\newcommand\riemann[2]{R^{#1}_{\hphantom{#1} #2}}
\newcommand\christoffel[2]{\Gamma^{#1}_{\hphantom{#1} #2}}

\usepackage{calrsfs}

\usepackage{xcolor}

\usepackage{amsthm}
\newtheorem{result}{Result}
\newtheorem*{corollary*}{Corollary}

\newcommand{\myskip}{\hspace{9pt}}

\begin{document}

\title{Is gravitational collapse possible in $f(R)$ gravity?}

\author{Adri\'{a}n Casado-Turri\'{o}n}
    \affiliation{Departamento de F\'{i}sica Te\'{o}rica and Instituto IPARCOS,  Universidad Complutense, 28040 Madrid, Spain}
\author{\'{A}lvaro de la Cruz-Dombriz}
    \affiliation{ Departamento de F\'{i}sica Fundamental, Universidad de Salamanca, 37008 Salamanca, Spain}
    \affiliation{Cosmology and Gravity Group, Department of Mathematics and Applied Mathematics, University of Cape Town, Rondebosch 7700, Cape Town, South Africa}
\author{Antonio Dobado}
    \affiliation{Departamento de F\'{i}sica Te\'{o}rica and Instituto IPARCOS,  Universidad Complutense, 28040 Madrid, Spain}

\date{\today}

\begin{abstract}
Gravitational collapse is still poorly understood in the context of $f(R)$ theories of gravity, since the Oppenheimer-Snyder model is incompatible with their junction conditions. In this work, we will present a systematic approach to the problem. Starting with a thorough analysis of how the Oppenheimer-Snyder construction should be generalised to fit within metric $f(R)$ gravity, we shall subsequently proceed to explore the existence of novel exterior solutions compatible with physically viable interiors. Our formalism has allowed us to show that some paradigmatic vacuum metrics cannot represent spacetime outside a collapsing dust star in metric $f(R)$ gravity. Moreover, using the junction conditions, we have found a novel vacuole solution of a large class of $f(R)$ models, whose exterior spacetime is documented here for the first time in the literature as well. Finally, we also report the previously unnoticed fact that the Oppenheimer-Snyder model of gravitational collapse is incompatible with the junction conditions of the Palatini formulation of $f(R)$ gravity.
\end{abstract}

\maketitle

\section{Introduction}
\label{Introduction}

$f(R)$ theories of gravity \cite{Sotiriou:2008rp,DeFelice:2010aj} are among the simplest possible extensions of General Relativity (GR). Fuelled by the discovery of the accelerated expansion of the Universe in 1998 \cite{SupernovaCosmologyProject:1998vns}, $f(R)$ theories became ubiquitous in cosmology \cite{Nojiri:2010wj,Nojiri:2017ncd} for two main reasons. First, within the $f(R)$ formalism, it is not necessary to include an \textit{ad hoc} dark-energy component in the stress-energy tensor of the Universe. Instead, the accelerated cosmic expansion arises naturally in $f(R)$ theories as a consequence of the modified gravitational dynamics. Second, it is straightforward to construct $f(R)$ models of gravity which are compatible not only with cosmological observations, but also with Solar System experiments and other local gravity constraints.

Despite their success in explaining cosmological observations, $f(R)$ theories of gravity have not proved to be equally fruitful when one attempts to describe compact-object dynamics. So far, only static stellar configurations have been studied within the $f(R)$ formalism, both in the relativistic \cite{AparicioResco:2016xcm,Astashenok:2017dpo} and non-relativistic cases ---for a review, see \cite{Olmo:2019flu}---. What is more, gravitational collapse is still poorly understood in the context of $f(R)$ gravity. Conversely, exact collapsing solutions in GR have been known from very early on. For example, the Oppenheimer-Snyder model \cite{Oppenheimer:1939ue}, describing the gravitational collapse of a uniform-density dust star, was conceived as early as 1939. The Oppenheimer-Snyder construction is particularly insightful in the sense that it is the simplest possible model of gravitational collapse. More realistic descriptions are all qualitatively similar to the Oppenheimer-Snyder picture, hence its importance.

When one attempts to describe the gravitational collapse of a uniform-density dust star in $f(R)$ gravity, most difficulties arise because the junction conditions of these theories differ from the renowned Darmois-Israel junction conditions of GR \cite{Darmois,Israel:1966rt}. In particular, junction conditions in $f(R)$ gravity put tighter bounds than those of GR, both in the Palatini \cite{Olmo:2020fri} and metric \cite{Deruelle:2007pt,Senovilla:2013vra} formalisms. On the one hand, any construction involving a dust star interior and a vacuum exterior is impossible on Palatini $f(R)$ gravity, as per its junction conditions.\footnote{To the best of our knowledge, the fact that dust stars are incompatible with the junction conditions of Palatini $f(R)$ gravity presented in \cite{Olmo:2020fri} has not been pointed out on any previous works, even though it is a straightforward consequence of such junction conditions. For the sake of completeness, we provide a discussion of this fact in Appendix \ref{Appendix Palatini}.} On the other hand, dust stars are not incompatible with the junction conditions of metric $f(R)$ gravity \textit{a priori}. However, progress towards a simple account of gravitational collapse within the metric formalism has been further hindered by the necessary non-triviality of the exterior spacetime, which is not known in principle \cite{Bueno:2017sui}. Because of this, previous works in the literature did not take into account the junction conditions nor the exterior, and instead focused on determining the evolution of the interior spacetime for various equations of state \cite{Cembranos:2012fd,Astashenok:2018bol}.

For all these reasons, in this work we shall endeavour to take a first step towards shedding some light on the issue of gravitational collapse in metric $f(R)$ gravity. We will consider the collapse of a spherically symmetric, uniform-density dust star under its own gravitational pull, with the purpose of extracting as much information as possible from the relevant junction conditions. In particular, we are able to obtain some no-go results which severely constrain the form of the exterior metric. As a byproduct, we find a previously undiscovered static vacuole solution of a large class of $f(R)$ theories of gravity. Remarkably, this novel solution is not a solution of GR. Moreover, as far as we are aware, the vacuole is one of the very few known glued spacetimes which satisfies all the relevant junction conditions of metric $f(R)$. It is therefore a highly non-trivial solution, despite its simple appearance. Our solution also has a vanishing Ricci scalar. The existence of such non-trivial solutions with constant scalar curvature in metric $f(R)$ gravity has been known for a long time \cite{delaCruz-Dombriz:2009pzc,Nzioki:2009av,Calza:2018ohl}. However, very few explicit examples can be found in the literature.

Our work is further motivated because of the existing connection between gravitational collapse in metric $f(R)$ gravity and the black-hole no-hair theorems (NHTs) \cite{Bueno:2017sui}. Metric $f(R)$ gravity is dynamically equivalent to a scalar-tensor theory \cite{Sotiriou:2008rp,DeFelice:2010aj}. It is well known that the NHTs hold in $f(R)$ theories whose additional scalar degree of freedom satisfies some technical (but still general and easily achievable) conditions \cite{Sotiriou:2011dz}. The NHTs guarantee that the only stationary, linearly-stable black holes resulting from gravitational collapse are those of GR, i.e.~they belong to the Kerr family. In particular, this implies that a non-rotating star should collapse into a Schwarzschild black hole in theories satisfying the NHTs. However, the resulting Schwarzschild spacetime would have trivial (i.e.~constant) scalar hair, while the spacetime outside the collapsing star is hairy ---as originally pointed out in \cite{Bueno:2017sui}, and discussed in Section \ref{intandext}---. Consequently, the scalar field should disappear dynamically as the star collapses. Hence, a detailed account of the process of gravitational collapse in metric $f(R)$ gravity should shed light on the mechanism by which a star could dispose of its originally non-trivial scalar hair.

\subsection{How this work is organised}

The article shall be organised as follows. In Section \ref{Section2} we present the rudiments for the collapse of spherical dust configurations in the context of metric $f(R)$ models of gravity. Therein, in Section \ref{OS incompatible with f(R)}, we shall revisit the incompatibility of the Oppenheimer-Snyder model with these theories. In Section \ref{intandext}, we provide the interior spacetime ---i.e.~a Friedmann-Lema\^itre-Robertson-Walker (FLRW)-like metric--- as well as the hypotheses for the exterior metric. In \ref{SystematicApproach}, we sketch the systematic approach to follow in the upcoming sections. Then, in Section \ref{junction conditions f(R)}, we shall briefly make the specific form of junctions conditions explicit when the matching of the interior and exterior spacetimes under consideration is imposed.

Subsequently, Sections \ref{ruling out} and \ref{staticsol} constitute the core of this investigation, so the busy reader is encouraged to focus on them. Section \ref{ruling out} presents a series of of five seminal results ---plus one corollary--- constraining the viable exterior spacetimes which can be smoothly connected with dust FLRW interiors. Proofs of such results appear in  Section \ref{Proofs Results 1 2 Corollary} and Section \ref{Proofs Results 3 4 5}. Section \ref{staticsol} is then devoted to the study of a novel vacuole solution constructed with a previously unknown static exterior and a Minkowski interior (which is obtained from the usual FLRW-like interior by neglecting gravitational collapse). The main features of this solution are studied here. Finally, we conclude our research with Section \ref{Conclusions}, where conclusions and future work prospects are summarised.

Most detailed calculations of some key issues covered in this article have been relegated to the appendices. Appendix \ref{Appendix A - Oppenheimer-Snyder} summarises the usual Oppenheimer-Snyder collapse model, such that differences can be spotted more easily when the underlying theory of gravity moves from GR to non-linear metric $f(R)$ scenarios. Appendix \ref{ArealRadiusJCAppendix} resorts to the so-called `areal-radius' coordinates to present in detail the $f(R)$ junction conditions in this system of coordinates, whereas Appendix \ref{NOTArealRadiusJCAppendix} presents such conditions without resorting to `areal-radius' coordinates. Finally, Appendix \ref{Appendix Palatini} briefly shows how the Oppenheimer-Snyder collapse model is unfeasible in the Palatini formulation of $f(R)$ gravity since dust stars are incompatible with the junction conditions presented in \cite{Olmo:2020fri}.

\subsection{Some minor technicalities}

As is widely known, the action of metric $f(R)$ gravity reads\footnote{Our sign convention shall be the one denoted as $(-,-,+)$ by Misner, Thorne and Wheeler \cite{Misner:1973prb}: the metric signature will be $(+,-,-,-)$, the Riemann tensor will be defined as $\riemann{\rho}{\sigma\mu\nu}=\partial_\nu\christoffel{\rho}{\sigma\mu}+\christoffel{\rho}{\lambda\nu}\christoffel{\lambda}{\sigma\mu}-(\mu\leftrightarrow\nu)$, the Ricci tensor will be given by $R_{\mu\nu}=\riemann{\lambda}{\mu\lambda\nu}$, and the Einstein field equations will read $G_{\mu\nu}=-\kappa T_{\mu\nu}$, with $\kappa\equiv 8\pi G$ ($c=1$). Also, in the following, sub-indices $t$ and $r$ denote differentiation with respect to coordinates $t$ and $r$, respectively. 
}
\begin{equation} \label{f(R) action}
    S=\dfrac{1}{2\kappa}\int\dif^4 x\,\sqrt{-g}\,f(R)+S_\mathrm{matter}.
\end{equation}
Its associated equations of motion are
\begin{equation} \label{f(R) EOM}
    f_R(R) R_{\mu\nu}-\dfrac{f(R)}{2}g_{\mu\nu}+\mathcal{D}_{\mu\nu}f_R(R)=-\kappa T_{\mu\nu},
\end{equation}
where $f_R(R)\equiv\dif f/\dif R$ and $\mathcal{D}_{\mu\nu}\equiv\nabla_\mu\nabla_\nu-g_{\mu\nu}\square$.

Hereafter, we will consistently refer to $f(R)$ theories satisfying $f_{RR}(R)\equiv\dif^2 f/\dif R^2\neq 0$ as `non-linear $f(R)$ gravity', in order to clearly differentiate them from `linear $f(R)$ gravity', i.e.~$f(R)=R-2\Lambda$ or, in other words, GR plus a cosmological constant.

Lastly, in this work we shall be interested in the smooth matching of two spacetimes, $\mathcal{M}^-$ and $\mathcal{M}^+$ (henceforth referred to as the `interior spacetime' and the `exterior spacetime', respectively) across a fixed time-like hypersurface. Because this hypersurface will correspond to a stellar surface, it shall be denoted as $\Sigma_*$. We assume that $\Sigma_*$ is endowed with intrinsic coordinates $y^a$ ($a=1,2,3$) and that, on either side of the boundary, there exist coordinates $x_\pm^\mu$ ($\mu=0,1,2,3$) that cover $\Sigma_*$.\footnote{In general, given a certain quantity $Q$, $Q^\pm$ shall refer to the values $Q$ takes in $\mathcal{M}^\pm$, respectively. Moreover, $Q_*^\pm$ will denote the values $Q$ takes at $\Sigma_*$ as $\Sigma_*$ is approached from $\mathcal{M}^\pm$, respectively. Finally, we define the jump of quantity $Q$ across the stellar surface $\Sigma_*$ as $[Q]\equiv Q_*^+-Q_*^-$. For example, if $[Q]=0$, then quantity $Q$ is obviously continuous across $\Sigma_*$.}

\section{Collapsing dust stars in non-linear metric $f(R)$ gravity}
\label{Section2}

\subsection{Revisiting the incompatibility of Oppenheimer-Snyder collapse in non-linear metric $f(R)$ gravity} \label{OS incompatible with f(R)}

Given its importance for our understanding of gravitational collapse, it is natural to wonder whether the Oppenheimer-Snyder construction, as described in the Appendix \ref{Appendix A - Oppenheimer-Snyder}, is a proper matched solution of non-linear $f(R)$ theories of gravity in the metric formalism. The answer is that it is not, as carefully shown in \cite{Senovilla:2013vra}, because the junction conditions of $f(R)$ gravity are different from those of GR \cite{Deruelle:2007pt,Senovilla:2013vra}. The modified junction conditions also imply that a larger class of collapse models are incompatible with metric $f(R)$ gravity as well \cite{Senovilla:2013vra,Goswami:2014lxa}.

Besides the renowned Darmois-Israel junction conditions of GR ---$[h_{ab}]=0$ and $[K_{ab}]=0$---, non-linear metric $f(R)$ gravity requires two additional constraints to be satisfied so as to make a given matching possible. In particular,
\begin{itemize}
    \item \textbf{Third junction condition}: the continuity of the Ricci scalar at $\Sigma_*$, i.e.~$[R]=0$, and
    \item \textbf{Fourth junction condition}: the continuity of the normal derivative of the Ricci scalar at $\Sigma_*$, i.e.~$[n^\mu\partial_\mu R]=0$.
\end{itemize}
The existence of these two supplementary junction conditions in metric $f(R)$ gravity is due to the fact that these theories propagate an additional scalar degree of freedom in comparison with GR, and that this scalar mode is intimately related to the Ricci scalar ---for further information, see, for instance, reference \cite{Sotiriou:2008rp}---.

Let us now comment on how the incompatibility of Oppenheimer-Snyder collapse arises in $f(R)$ gravity.\footnote{In the original proof of the incompatibility \cite{Senovilla:2013vra}, the author considers several known glued solutions of GR ---including Oppenheimer-Snyder---, and then determines whether they are solutions of $f(R)$ gravity as well. His derivation consists of a proof by \textit{reductio ad absurdum}: the Oppenheimer-Snyder construction is assumed to be a solution of $f(R)$ gravity, and then the author shows that this implies that the dust star has vanishing energy density everywhere. Herein, we shall follow a different, although equivalent, approach.} It is well known that both the interior FLRW metric \eqref{FLRW} and the exterior Schwarzschild spacetime are solutions of $f(R)$ gravity. In the case of FLRW, only the dependence of $a(\tau)$ in $\tau$ gets modified \cite{Cembranos:2012fd}, as we shall see later, while Schwarzschild is a solution of every $f(R)$ model satisfying $f(0)=0$ \cite{delaCruz-Dombriz:2009pzc}. If one attempted to glue these two spacetimes at the stellar surface, the third junction condition $[R]=0$ would then require
\begin{equation} \label{3 OS f(R)}
    R^-_*=6\left(\dfrac{\dot{a}^2+k}{a^2}+\dfrac{\ddot{a}}{a}\right)=R^+_*=0.
\end{equation}
This is an ordinary differential equation for the scale factor $a(\tau)$ which may be integrated using the standard initial conditions $a(0)=1$ and $\dot{a}(0)=0$, yielding
\begin{equation} \label{constantRa}
    a(\tau)=\sqrt{1-k\tau^2}.
\end{equation}
Therefore, the third junction condition ---which is exclusive to non-linear metric $f(R)$ gravity--- \textit{fixes} the scale factor of the interior spacetime to be given by equation \eqref{constantRa}, instead of the cycloid equation \eqref{cycloid} one has in GR. Moreover, as explained in Appendices \ref{Appendix A - Oppenheimer-Snyder} and \ref{ArealRadiusJCAppendix}, fixing $a(\tau)$ amounts to fixing the evolution of $r_*(\tau)$ and $t_*(\tau)$ ---i.e.~of the stellar surface \eqref{stellarsurface out r}---, by virtue of the first junction condition.

At first sight, \eqref{constantRa} seems to be an appropriate replacement of the cycloid-like scale factor obtained from expression \eqref{cycloid}; at least, it shares a number of its most distinguishing features. Indeed, an interior FLRW spacetime with a scale factor given by expression \eqref{constantRa} would still be \textit{dynamic}, despite having a vanishing Ricci scalar. Moreover, if $k>0$, the star would collapse to zero proper volume in finite proper time, just as in the Oppenheimer-Snyder model. And, remarkably, \textit{all four} junction conditions would be satisfied by construction.\footnote{If $R^+=R^-=0$, then $[n^\mu\partial_\mu R]=0$, and the fourth junction condition is satisfied automatically.} However, we must bear in mind that equation \eqref{3 OS f(R)} is simply an additional \textit{constraint}, in the sense that it is a relationship between metrics. This relationship may or may not be compatible with the equations of motion of $f(R)$ gravity. In other words, the third junction condition requires the scale factor to be given by \eqref{constantRa}; this, in turn, implies that the matching between the interior FLRW spacetime and the exterior Schwarzschild metric can only occur in those $f(R)$ theories whose equations of motion give rise to a scale factor which evolves in $\tau$ according to expression \eqref{constantRa}. As we shall prove in Section \ref{ruling out}, an interior FLRW spacetime whose scale factor is given by \eqref{constantRa} does not solve the equations of motion of $f(R)$ gravity for any choice of function $f$. 
In consequence, Oppenheimer-Snyder collapse is not possible in metric $f(R)$ gravity, as anticipated. Furthermore, we clearly see that the incompatibility arises because we have insisted that the exterior spacetime is Schwarzschild. Thus, we are led to conclude that, in non-linear $f(R)$ gravity, the exterior must be a different, more general spacetime.

\subsection{Interior and exterior metrics} \label{intandext}

Since the Oppenheimer-Snyder model of collapse is no longer a valid matched solution of non-linear $f(R)$ gravity in the metric formalism, one needs to reconsider whether metrics \eqref{FLRW} and \eqref{Schwarzschild} correctly describe the interior and the exterior of the collapsing uniform-density star (respectively) in these theories. We shall assume that the interior stress-energy tensor corresponds to dust in the Jordan frame representation of the theory \eqref{f(R) action} and not in the conformally-related Einstein frame.

As shown in \cite{Cembranos:2012fd}, the spacetime corresponding to a spherically-symmetric, uniform-density distribution of dust in any $f(R)$ theory of gravity is still a portion of FLRW spacetime \eqref{FLRW}. Thus, the $f(R)$ field equations do not change the form of the interior metric; however, they do alter the scale factor dynamics. More precisely, the equation for $a(\tau)$ now becomes
\begin{equation} \label{ap2 f(R)}
    \dot{a}^2=-k+\dfrac{1}{f_R^-}\left(\dfrac{\kappa\rho_0}{6a}+\dfrac{a^2}{2}\ddot{f}_R^-+\dfrac{\dot{a}a}{2}\dot{f}_R^-+\dfrac{a^2}{6}f^-\right),
\end{equation}
which reduces to the cycloid equation \eqref{cycloid} of GR when $f(R)\equiv R$, as expected.\footnote{Throughout the text, we employ the usual notation $f^-\equiv f(R^-)$, $f_R^-\equiv f_R(R^-)$.
Analogously, we will also use $f^+\equiv f(R^+)$ and $f_R^+\equiv f_R(R^+)$ later on.} Assuming the usual initial conditions $a(0)=1$ and $\dot{a}(0)=0$, the expression for $k$ is modified accordingly,
\begin{equation} \label{k f(R)}
    k=\dfrac{\kappa\rho_0+3\ddot{f}_{R0}^- +f_0^-}{6f_{R0}^-},
\end{equation}
where $f_0^-\equiv f(R_0^-)$, $f_{R0}^-\equiv f_R(R_0^-)$ and $R_0^-\equiv R^-(0)$.

As mentioned in Section \ref{OS incompatible with f(R)}, the spacetime outside the star cannot be Schwarzschild. This assertion is also supported by a theorem in \cite{Bueno:2017sui}, which states that only in theories that exclusively propagate a traceless and massless graviton \textit{in vacuo} the gravitational field outside a spherically symmetric mass distribution can be represented by metrics of the form
\begin{equation} \label{singlefunction}
    \dif s_+^2=A(r)\,\dif t^2-A^{-1}(r)\,\dif r^2-r^2\dif\Omega_{(2)}^2,
\end{equation}
of which Schwarzschild is a particular example. As mentioned before, it is well known that, apart from the usual massless and traceless graviton, non-linear metric $f(R)$ theories propagate an additional scalar degree of freedom, known as the scalaron \cite{Sotiriou:2008rp,DeFelice:2010aj}. In the Einstein-frame representation of the theory, the scalaron is given by
\begin{equation} \label{scalaron}
    \phi=\sqrt{\dfrac{3}{2\kappa}}\ln f_R(R).
\end{equation}
Thus, the assertion in \cite{Bueno:2017sui} guarantees that, in non-linear metric $f(R)$ gravity, Schwarzschild can only exist as a black hole, but not as an exterior spacetime matching interior matter distributions whatsoever, even though Schwarzschild remains as a vacuum solution of the field equations. One can intuitively understand this result by noticing that the scalaron \eqref{scalaron} will be excited provided that there is matter somewhere in the spacetime, since (as widely known) the former couples to the trace of the stress-energy tensor. However, equation \eqref{scalaron} reveals that Schwarzschild spacetime only supports a trivial (i.e.~constant) scalar field.\footnote{Herein $f_R(0)\neq 0$ is assumed; otherwise, the corresponding scalar field would not be properly defined.}

Since Birkhoff's theorem does not hold in $f(R)$ theories of gravity,\footnote{The strongest result in this respect is that the Schwarzschild spacetime is the only static, spherically symmetric solution \textit{with vanishing Ricci scalar} in $f(R)$ theories satisfying $f(0)=0$ and $f_R(0)\neq 0$, cf.~reference \cite{Nzioki:2009av}, but this does not exclude the possibility that there exist other exterior vacuum solutions with a non-constant Ricci scalar even in these theories.} virtually any spherically symmetric vacuum line element could match FLRW interior uniform-density dust solutions. As a result, we are obliged to resort to the full set of junction conditions in order to determine which exterior metric is the correct one. Without further guidance, it may seem that the problem basically consists in `looking for a needle in a haystack'; in spite of this, the junction conditions of $f(R)$ gravity impose strong constraints on the exterior of the uniform-density dust star, as we shall show in Section \ref{ruling out}.

\subsection{A systematic approach for junction conditions}
\label{SystematicApproach}

Given that we do not know \textit{a priori} what the spacetime outside a collapsing uniform-density dust star is in $f(R)$ gravity ---in fact, we can only assume that it is spherically symmetric, by analogy with the interior spacetime \eqref{FLRW}---, there are essentially two ways of approaching the problem:
\begin{itemize}
    \item either one attempts to infer properties of the exterior metric using the junction conditions, choosing for the exterior spacetime the most general spherically symmetric line element ---or a less general, but still spherically symmetric ansatz---, or
    \item one tries to show whether a known spherically-symmetric vacuum solution of $f(R)$ gravity matches the interior FLRW spacetime, by explicitly checking that all four junction conditions are satisfied simultaneously.
\end{itemize}
As mentioned in Section \ref{Introduction}, $f(R)$ theories admit a variety of exterior vacuum solutions. Thus,  it is reasonable to use junction conditions in the aim of checking whether any known metric satisfies them. Nonetheless, junction conditions must be handled with special care when one tries to extract information about the exterior spacetime from them. Indeed, despite the simple form they might turn out to take, the most intuitive line of reasoning may present some loopholes, and apparent exceptions become possible, as we will see in the following.

Let us consider, for example, the assertion made in \cite{Goswami:2014lxa} claiming that, for any $f(R)$ gravity with a nonlinear function $f$, a dynamic homogeneous spacetime with non-constant Ricci scalar cannot be matched with a static spacetime across a fixed boundary. Therein, authors argued that this result follows from the third junction condition ---i.e.~$[R]=0$---, because $R^-_*$ would be a function of $\tau$ only, while $R^+_*$ would be $\tau$-independent. Thus, equating both quantities for all $\tau$ would be impossible.

However, one can exploit a hole in this reasoning so as to find a static exterior solution that can be matched with the interior FLRW spacetime. The key is to realise that what $[R]=0$ implies is that, if the interior solution has a Ricci scalar $R^-=R^-(\tau)\neq\const$, then the exterior solution cannot be static \textit{with respect to time coordinate} $\tau$ ---i.e.~it cannot have a time-like Killing vector field $\partial_\tau$---, but this does not prevent the exterior from being static with respect to a different time coordinate $t$ ---i.e.~it could have a different time-like Killing vector field $\partial_t$---. In order to illustrate this, let us consider an interior FLRW spacetime and an exterior spacetime which is static with respect to some time coordinate $t$. Then it is always possible to choose `areal-radius' coordinates $(t,r,\theta,\varphi)$ in which the latter line element takes the form
\begin{equation}
\label{MetricStaticAB}
    \dif s_+^2=A(r)\,\dif t^2-B(r)\,\dif r^2-r^2\dif\Omega_{(2)}^2.
\end{equation}
Let us also assume that the exterior metric has a non-constant Ricci scalar $R^+=R^+(r)$.\footnote{This is perfectly reasonable even for a vacuum solution of non-linear $f(R)$ gravity because, \textit{in vacuo}, the trace of the $f(R)$ equations of motion is a differential equation for the scalar curvature ---rather than algebraic relation, as in GR, where $T^+=0$ necessarily implies $R^+=0$---.} As shown in the case of Schwarzschild in Appendix \ref{Appendix A - Oppenheimer-Snyder} ---see for instance equation \eqref{stellarsurface out r}---, the radial coordinate of this spacetime will become a function of $\tau$ upon changing to interior coordinates $(\tau,\chi,\theta,\varphi)$. Therefore, $R^+_*$ will generically depend upon $\tau$ even though the exterior metric is static, and thus the matching with a $\tau$-dependent interior scalar curvature (such as that of FLRW spacetime) could still be possible.

Nonetheless, we shall see in Section \ref{ruling out} that a thorough treatment of the junction conditions (including their compatibility with the field equations) rules out all static exteriors, and thus the conclusions in \cite{Goswami:2014lxa} remain valid, although resorting to reasons different from those given in that reference.

Finally, even though a variety of simple ans\"{a}tze for the exterior metric are not ruled out by the junction conditions ---as will be shown in Section \ref{ruling out}---, another important point to be considered is that one must always make sure that the constraints imposed by the junction conditions are compatible with the $f(R)$ equations of motion. Recall, for instance, the discussion of Section \ref{OS incompatible with f(R)}. As explained there, the third junction condition led to a result ---equation \eqref{constantRa}--- which was incompatible with the equations of motion of $f(R)$ gravity. We thus concluded that the matching was impossible. Analogous scenarios will appear when considering exterior spacetimes in Section \ref{ruling out} below.

\section{Junction conditions in non-linear metric $f(R)$ gravity} \label{junction conditions f(R)}

As per the discussion in the previous sections, the only assumption one can make on the exterior spacetime is that it must be spherically symmetric. Consequently, our starting point will be the most general spherically symmetric line element. We aim at determining which conditions such a spacetime should satisfy so as to match an interior FLRW-like metric.

In order to establish the junction conditions, first a coordinate system in which the exterior metric is expressed needs to be chosen. The most natural one is the coordinate system $(t,r,\theta,\varphi)$ in which the line element is of the form
\begin{equation} \label{sphericallysymmetric}
    \dif s^2_+=A(t,r)\,\dif t^2 -B(t,r)\,\dif r^2-r^2\,\dif \Omega_{(2)}^2,
\end{equation}
i.e.~$r$ is the `areal radius' in the sense that spherical time-like hypersurfaces centered at the origin have proper area $4\pi r^2$. The stellar surface is still given by \eqref{stellarsurface out r} using these coordinates. `Areal-radius' coordinates $(t,r,\theta,\varphi)$ are specially convenient since junction conditions remain very similar to those of Oppenheimer-Snyder collapse. Moreover, each of the four junction conditions to follow essentially retains its interpretation given in Appendix \ref{Appendix A - Oppenheimer-Snyder}. For the thorough derivation of the junctions conditions appearing immediately below, we refer the reader to Appendix \ref{ArealRadiusJCAppendix}.

Using `areal-radius' coordinates, there are six independent equations coming from the four junction conditions. Two of these equations,
\begin{gather}
    r_*=a\chi_*, \label{1.1 f(R)} \\
    A_*\dot{t}_*^2-B_*\dot{r}_*^2=1, \label{1.2 f(R)}
\end{gather}
are obtained by requiring $[h_{ab}]=0$, i.e.~by imposing the first junction condition. Just as in Oppenheimer-Snyder collapse, \eqref{1.1 f(R)} and \eqref{1.2 f(R)} unequivocally determine the evolution of the matching surface $\Sigma_*$. On the one hand, one can clearly see that equation \eqref{1.1 f(R)} reveals that the stellar areal radius $r_*(\tau)$ is simply proportional to $a(\tau)$. On the other hand, equation \eqref{1.2 f(R)} is a first-order ordinary differential equation for $t_*(\tau)$,\footnote{Since $r_*(\tau)\propto a(\tau)$ as per the first junction condition \eqref{1.1 f(R)}, the $n$-th derivative of $r_*$ with respect to $\tau$ is proportional to the $n$-th derivative of $a$. One can determine $a(\tau)$ by solving \eqref{ap2 f(R)}; once $a(\tau)$ is known, all of its derivatives can be obtained in turn. Therefore, of all the functions of $\tau$ appearing on equation \eqref{1.2 f(R)}, the only one which is unknown ---unless we solve \eqref{1.2 f(R)}--- is $t_*(\tau)$. As a result, junction condition \eqref{1.2 f(R)} is always a differential equation for $t_*(\tau)$.} and thus its solution is unique given an initial condition. Thus, these two equations allow us to know whether the star either collapses, expands or bounces, depending on the specific dynamics for the scale factor $a(\tau)$ associated to a given choice for the $f(R)$ model. Therefore, the interpretation of the first junction condition remains unchanged with respect to GR in $f(R)$ gravity.

The second pair of equations,
\begin{gather}
    \dot{\beta}=\dfrac{A_{t*}\dot{t}_*^2-B_{t*}\dot{r}_*^2}{2}, \label{2.1 f(R)} \\
    \beta=\beta_0\sqrt{A_* B_*}, \label{2.2 f(R)}
\end{gather}
is obtained by requiring $[K_{ab}]=0$, i.e.~by imposing the second junction condition. Here, 
\begin{equation} \label{betadeff(R)}
    \beta\equiv A_*\,\dot{t}_*=\sqrt{A_*+A_*B_*\dot{r}_*^2},
\end{equation}
and
\begin{equation} \label{beta0 f(R)}
    \beta_0\equiv\sqrt{1-k\chi_*^2}.
\end{equation}
In Oppenheimer-Snyder collapse (in which $B_*=1/A_*$ and $A_{t*}=B_{t*}=0$), expressions \eqref{2.1 f(R)} and \eqref{2.2 f(R)} are satisfied simultaneously, and thus they actually provide only one condition. In the general case, however, \eqref{2.1 f(R)} and \eqref{2.2 f(R)} are distinct equations. What remains unchanged is the interpretation of the second junction condition, namely that equations \eqref{2.1 f(R)} and \eqref{2.2 f(R)} should provide a relationship between the parameters of the interior and exterior spacetimes, if the constraints imposed by the other junction conditions are also taken into account.

Finally, the remaining two junction conditions, which are exclusive to $f(R)$ gravity, are
\begin{gather}
    R^+_*=6\left(\dfrac{\dot{r}_*^2+k\chi_*^2}{r_*^2}+\dfrac{\ddot{r}_*}{r_*}\right), \label{3 f(R)} \\
    \dfrac{\dot{r}_*}{A_*}R_{r*}^++\dfrac{\dot{t}_*}{B_*}R_{t*}^+=0. \label{4 f(R)}
\end{gather}
These two additional constraints arise from imposing $[R]=0$ and $[n^\mu\partial_\mu R]=0$, respectively.

To sum up, the relevant junction conditions between the interior FLRW spacetime \eqref{FLRW} and the spherically symmetric exterior \eqref{sphericallysymmetric} in non-linear $f(R)$ gravity are equations \eqref{1.1 f(R)}--\eqref{2.2 f(R)}, \eqref{3 f(R)} and \eqref{4 f(R)}, together with the definitions of $\beta$ and $\beta_0$ given by equations \eqref{betadeff(R)} and \eqref{beta0 f(R)}, respectively.

At this stage, we must note that the behaviour of the stellar surface is already fixed by the first junction condition alone ---i.e.~by equations \eqref{1.1 f(R)} and \eqref{1.2 f(R)}---. Thus, in principle, expressions \eqref{2.1 f(R)}, \eqref{2.2 f(R)}, \eqref{3 f(R)} and \eqref{4 f(R)} should only contribute to parameter determination. A similar situation occurs in Oppenheimer-Snyder collapse: the second junction condition ultimately provides the relationship \eqref{M OS} between $M$, $k\propto\rho_0$ and $\chi_*$. 

Finally, we shall stress that the procedure by which the junction conditions are obtained is fully coordinate-dependent. Accordingly, any coordinate change in expression \eqref{sphericallysymmetric} would require the junction conditions to be derived again. In the context of non-static spacetimes, this introduces some additional difficulties. For example, if the exterior spacetime is naturally given in a coordinate system different than $(t,r,\theta,\varphi)$, the change to `areal-radius' coordinates might be very difficult ---if not impossible--- to perform analytically. Moreover, the junction conditions may become harder to implement if they are derived in a coordinate system different from $(t,r,\theta,\varphi)$. For example, it is always possible to choose coordinates $(\eta,\xi,\theta,\varphi)$ in which the spherically symmetric exterior line element can be expressed as
\begin{equation}
    \dif s^2_+=C(\eta,\xi)\,\dif \eta^2 -D(\eta,\xi)\,\dif\xi^2-r^2(\eta,\xi)\,\dif \Omega_{(2)}^2.
\end{equation}
Some of the most renowned non-trivial vacuum solutions of metric $f(R)$ gravity, such as the so-called Clifton II spacetime \cite{Clifton:2006ug} are naturally given in this form, hence its importance. The derivation of the junction conditions using these coordinates, as well as a discussion on their convoluted interpretation, is presented in Appendix \ref{NOTArealRadiusJCAppendix}.

\section{Exterior metrics forbidden by the junction conditions} \label{ruling out}

In this section, we intend to obtain constraints on the exterior spacetime by means of the junction conditions of $f(R)$ gravity, that is to say, equations \eqref{1.1 f(R)}--\eqref{2.2 f(R)}, \eqref{3 f(R)} and \eqref{4 f(R)} presented in the previous section. The specific form of these conditions is valid provided that the exterior line element is given by expression \eqref{sphericallysymmetric}, i.e.~when one uses the ideally suited `areal-radius' coordinates $(t,r,\theta,\varphi)$.

Thus, we aim at building a system of differential equations for functions $A(t,r)$ and $B(t,r)$ in \eqref{sphericallysymmetric} out of the junction conditions. One can immediately notice that equations \eqref{1.1 f(R)}--\eqref{2.2 f(R)}, \eqref{3 f(R)} and \eqref{4 f(R)} are not given exclusively in terms $A_*$ and $B_*$; they also depend on $t_*$, $r_*$ and their $\tau$-derivatives. Accordingly, the first step will then be to express $\dot{t}_*$ and $\dot{r}_*$ in terms of $A_*$ and $B_*$. This can be easily achieved by combining expressions \eqref{1.2 f(R)}, \eqref{2.2 f(R)} and \eqref{betadeff(R)}. One then obtains
\begin{gather}
    \dot{t}_*=\beta_0\sqrt{\dfrac{B_*}{A_*}}, \label{tp} \\
    \dot{r}_*=\sqrt{\beta_0^2-\dfrac{1}{B_*}} \label{rp}.
\end{gather}
Expressions \eqref{tp} and \eqref{rp} may now be substituted back in equations \eqref{2.1 f(R)}, \eqref{2.2 f(R)}, \eqref{3 f(R)} and \eqref{4 f(R)}. For example, condition \eqref{2.1 f(R)} becomes
\begin{equation}
    \dot{\beta}=\dfrac{1}{2}\left[A_{t*} \dfrac{\beta_0^2 B_*}{A_*}-B_{t*} \left(\beta_0^2-\dfrac{1}{B_*}\right)\right],
\end{equation}
while equation \eqref{4 f(R)} turns into
\begin{equation} \label{inserted 4 f(R)}
    \sqrt{\beta_0^2-\dfrac{1}{B_*}}R_{r*}^++\beta_0\sqrt{\dfrac{A_*}{B_*}}R_{t*}^+=0.
\end{equation}

Proceeding in a similar fashion, the whole system of equations is re-expressed in such a way that only $A_*$, $B_*$ and their derivatives (with respect to $t$ and $r$) appear.
In principle, these equations hold on the stellar surface only, i.e.~on $t=t_*(\tau)$ and $r=r_*(\tau)$. However, in order to obtain differential equations involving $A(t,r)$ and $B(t,r)$, we can require the system to be satisfied for all $(t,r)$. For example, instead of \eqref{inserted 4 f(R)}, we may demand
\begin{equation} \label{all t and r 4 f(R)}
    \sqrt{\beta_0^2-\dfrac{1}{B}}R_{r}^++\beta_0\sqrt{\dfrac{A}{B}}R_{t}^+=0
\end{equation}
to be satisfied. Certainly, should this equation hold, then \eqref{inserted 4 f(R)} will hold as well. Moreover, requiring the junction conditions to be satisfied for all $t$ and all $r$ is indeed a reasonable assumption. If the star ends up collapsing, then $r_*$ will evolve continuously from its initial value $r_*(0)=\chi_*$ (which is arbitrary\footnote{In general, the surface of a star is chosen to be any spherical surface in which the stellar pressure $p$ vanishes. Therefore, in realistic stars (whose interior pressure is non-zero), there might be upper or lower bounds on the radius depending on the dynamics of $p$. However, in dust stars, the pressure vanishes identically everywhere in spacetime. Therefore, any spherical surface can be the initial stellar surface, and $r_*(0)=\chi_*$ is thus arbitrary.\label{chistar arbitrary}}) to zero. Similarly, $t_*(0)$ is also arbitrary, and one expects that the black hole resulting from gravitational collapse takes infinite exterior time $t$ to form ---this is the case, for example, in Oppenheimer-Snyder collapse \cite{Weinberg:1972kfs}---. As a result, assuming that $t_*(\tau)$ and $r_*(\tau)$ can take any values coordinates $t$ and $r$ can respectively take remains a well motivated hypothesis. Thus, it is sensible to require the junction conditions to hold for all allowed values of $t$ and $r$.

Demanding that junction conditions \eqref{1.1 f(R)}--\eqref{2.2 f(R)}, \eqref{3 f(R)} and \eqref{4 f(R)} hold for every $t$ and $r$, we are able to construct a system of differential equations, which includes \eqref{all t and r 4 f(R)},\footnote{Equation \eqref{all t and r 4 f(R)} is the most compact equation in the aforementioned system; the remaining ones are too long to be included in the bulk of the text.} to be satisfied by the exterior spacetime if it is to match the interior one, \eqref{FLRW}.
The intrinsic difficulty of solving this system for 
$A(t,r)$ and $B(t,r)$ ---or even of extracting any kind of information on the exterior spacetime from it--- renders the approach of specifying certain ans\"{a}tze for the exterior metric pragmatic. By proceeding this way, we have been able to establish the following no-go results:

\begin{result} \label{theoconstcurv}
    No exterior spacetime with constant scalar curvature ---either static or not--- can be smoothly matched to a uniform-density dust star interior in non-linear $f(R)$ gravity.
\end{result}

\begin{result} \label{theostatic}
    No static exterior spacetime can be smoothly matched to a uniform-density dust star interior in non-linear $f(R)$ gravity.
\end{result}

\begin{result} \label{theosinglefunction}
    No exterior metric of the form
    \begin{equation} \label{singlefunctionmetric}
        \dif s^2_+=A(t,r)\,\dif t^2-A^{-1}(t,r)\,\dif r^2-r^2\dif\Omega_{(2)}^2
    \end{equation}
    can be smoothly matched to a uniform-density dust star interior in non-linear $f(R)$ gravity.
\end{result}

\begin{result} \label{theoalmostANV}
    No exterior metric of the form
    \begin{equation}
    \label{eqnResult4}
        \dif s^2_+=U(t,r)\,V(r)\,\dif t^2-V^{-1}(r)\,\dif r^2-r^2\dif\Omega_{(2)}^2
    \end{equation}
    can be smoothly matched to a uniform-density dust star interior in non-linear $f(R)$ gravity.
\end{result}

\begin{result} \label{theoanothernonstatic}
    No exterior metric of the form
    \begin{equation}
     \label{eqnResult5}
        \dif s^2_+=U(r)\,\dif t^2-V(t)\,U^{-1}(r)\,\dif r^2-r^2\dif\Omega_{(2)}^2
    \end{equation}
    can be smoothly matched to a uniform-density dust star interior in non-linear $f(R)$ gravity.
\end{result}

Result \ref{theoconstcurv} above together with the fourth junction condition \eqref{all t and r 4 f(R)} also implies the following:

\begin{corollary*}
    No exterior spacetime of the form \eqref{sphericallysymmetric} can be smoothly matched to a uniform-density dust star interior in non-linear $f(R)$ gravity if its Ricci scalar $R^+$ is either a function of $t$ only or of $r$ only.
\end{corollary*}

In the following we will offer a proof of each of these statements.

\subsection{Proof of Result \ref{theoconstcurv}, its Corollary, and Result \ref{theostatic}}
\label{Proofs Results 1 2 Corollary}

Let us first consider a static exterior spacetime of the form \eqref{MetricStaticAB}: the starting hypothesis of Result \ref{theostatic}. With this particular choice for functions $A$ and $B$, junction condition \eqref{2.1 f(R)} reads $\dot{\beta}=0$, and thus we must require the $\tau$ derivative of \eqref{2.2 f(R)} to vanish. This yields
\begin{equation} \label{static beta dot}
    \sqrt{\beta_0^2-\dfrac{1}{B}}\,(A_r B+A B_r)=0,
\end{equation}
which is satisfied provided that
\begin{equation}
\label{Eq29}
    B(r)=\dfrac{1}{\beta_0^2}\myskip\text{or}\myskip A(r)\,B(r)=\const,
\end{equation}
depending on whether we demand either the square root or the parenthesis in \eqref{static beta dot} to vanish. The former choice is inconsistent with gravitational collapse, since it implies that $\dot{r}_*=0$ as per equation \eqref{rp}. Therefore, we shall only concentrate on the latter, i.e.~$B(r)=\const/A(r)$,\footnote{Nonetheless, by choosing the first option, $B(r)=1/\beta_0^2=\const$, we have been able to obtain a novel static, non-collapsing solution of $f(R)$ gravity, as we shall prove on Section \ref{staticsol}.} where the constant can always be set to 1 through a suitable redefinition of time coordinate $t$ ($\dif t\rightarrow\dif t/\const$). Junction condition \eqref{all t and r 4 f(R)} then becomes
\begin{equation}
    \sqrt{\beta_0^2-A}\,R_{r}^+=0.
\end{equation}
This equality is satisfied if either $A(r)=1/B(r)=\beta_0^2$ or $R^+=\const$ The former case must be discarded once again as explained above; therefore, the assumption of $R^+=\const$ renders junction condition \eqref{3 f(R)} as follows:
\begin{equation} \label{JC static constant curvature single function}
    R^+=\dfrac{6}{r^2}\left(1-A-\dfrac{r A_r}{2}\right),
\end{equation}
whose the right-hand side has been obtained by substituting \eqref{rp} on the right-hand side of \eqref{3 f(R)}. The general solution of \eqref{JC static constant curvature single function} is
\begin{equation} \label{mRN(A)dS}
    A(r)=1+\dfrac{Q^2}{r^2}-\dfrac{R^+ r^2}{12},
\end{equation}
$Q$ being an integration constant. This is the massless\footnote{We would like to highlight that a mass term $-2GM/r$ is present in the general solution of $R^+=\const$ for a static, spherically symmetric spacetime with $A(t,r)=1/B(t,r)=A(r)$. However, compliance with the third junction condition $[R]=0$ explicitly requires $M\neq0$. Therefore, the mass term, which was of paramount importance in standard Oppenheimer collapse, is absent from \eqref{mRN(A)dS} due to one of the novel junction conditions of non-linear metric $f(R)$ gravity. This is indeed a remarkable fact.} Reissner-Nordstr\"{o}m (Anti-)de Sitter spacetime, which is known to be a solution of any $f(R)$ theory coupled to an electromagnetic field \cite{delaCruz-Dombriz:2009pzc}.

Therefore, we have found that the only static and spherically symmetric solution of $f(R)$ gravity which satisfies junction conditions \eqref{2.1 f(R)}, \eqref{2.2 f(R)} and \eqref{3 f(R)} ---and also \eqref{4 f(R)}--- is \eqref{mRN(A)dS}. This solution possesses another crucial property: its Ricci scalar is constant. Therefore, if we are able to prove that constant-curvature solutions are incompatible with gravitational collapse in non-linear $f(R)$ gravity (Result \ref{theoconstcurv}), then will have also proved that the exterior cannot be static (Result \ref{theostatic}).

Consequently, let us now prove Result \ref{theoconstcurv}. Consider a spherically symmetric exterior spacetime with $R^+=\const$ This spacetime could either be static, such as \eqref{mRN(A)dS}, or non-static; our proof covers both situations. Junction conditions \eqref{1.1 f(R)} and \eqref{3 f(R)} would imply that the Ricci scalar of the interior FLRW spacetime must also be constant:
\begin{equation} \label{constantR+a}
   R^-= 6\left(\dfrac{\dot{a}^2+k}{a^2}+\dfrac{\ddot{a}}{a}\right)=R^+=\const
\end{equation}
In non-linear metric $f(R)$ gravity, it can be shown \cite{Cembranos:2012fd} that, for constant $R^-=R^+$,
\begin{equation}
    \dfrac{\ddot{a}}{a}=-\dfrac{2(\dot{a}^2+k)}{a^2}+\dfrac{f^+}{2f_{R}^+}.
\end{equation}
where $f^+\equiv f(R^+)$ and $f_{R}^+\equiv f_R(R^+)$. Substituting this expression in \eqref{constantR+a}, we find that
\begin{equation}
    -\dfrac{\dot{a}^2+k}{a^2}+\dfrac{f^+}{2f_{R}^+}=\dfrac{R^+}{6}.
\end{equation}
For constant $R^-=R^+$, equation \eqref{ap2 f(R)} would also require
\begin{equation}
    \dfrac{\dot{a}^2+k}{a^2}=\dfrac{1}{f_{R}^+}\left(\dfrac{\kappa\rho_0}{6a^3}+\dfrac{f^+}{6}\right),
\end{equation}
and we thus finally have that
\begin{equation}
    -\dfrac{\kappa\rho_0}{a^3}+2f^+=f_{R}^+ R^+,
\end{equation}
which can be reformulated as
\begin{equation} \label{constantaconstantR+}
    a^3(\tau)=\dfrac{\kappa\rho_0}{2f^+-f_{R}^+ R^+}=\const
\end{equation}
This result is incompatible with gravitational collapse, as per the first junction condition \eqref{1.1 f(R)}.\footnote{As anticipated in Section \ref{OS incompatible with f(R)}, equation \eqref{constantaconstantR+} ---evaluated at $R^+=0$--- is also incompatible with \eqref{constantRa}, which is obtained by direct integration of \eqref{3 OS f(R)} or \eqref{constantR+a} ---again evaluated at $R^+=0$---.} As a result, we have proved Result \ref{theoconstcurv}, i.e.~that no constant-curvature exterior solution ---either static or not--- can be matched to a collapsing dust star interior. $\blacksquare$

Since \eqref{mRN(A)dS} ---which is the only static solution satisfying the second, third and fourth junction conditions--- 
happens to have constant scalar curvature, it cannot be matched to the collapsing dust star interior due to Result \ref{theoconstcurv}, and thus we have also proved Result \ref{theostatic}. $\blacksquare$

Furthermore, because \eqref{mRN(A)dS} is also a `single-function' spacetime ---i.e.~of the form \eqref{singlefunction}---, we must stress that Result \ref{theostatic} is in agreement with the theorem in \cite{Bueno:2017sui} discussed back in Section \ref{intandext}.

Finally, the Corollary issued from  Result \ref{theoconstcurv} follows almost immediately from the fourth junction condition: if either $R^+_t$ or $R^+_r$ vanish, then equation \eqref{all t and r 4 f(R)} forces the other derivative ---$R^+_r$ or $R^+_t$, respectively--- to vanish as well. Consequently, the exterior solution would have a constant Ricci scalar. This is forbidden by Result \ref{theoconstcurv}. $\blacksquare$

To sum up, throughout Section \ref{Proofs Results 1 2 Corollary} we have found that if there exists a spherically symmetric exterior solution smoothly matching a dust star interior in $f(R)$ gravity, then such a solution must be non-static and have a non-constant Ricci scalar. Furthermore, its scalar curvature $R^+$ cannot depend only on either $t$ or $r$ when expressed in `areal-radius' coordinates.

\subsection{Proof of Results \ref{theosinglefunction}, \ref{theoalmostANV} and \ref{theoanothernonstatic}}
\label{Proofs Results 3 4 5}

Having discarded the possibility of having a static exterior spacetime, we shall now analyse one of the simplest non-static ans\"{a}tze one can possibly conceive: that given by expression \eqref{singlefunctionmetric}. However, it is almost immediate to show that the exterior metric cannot be of the form \eqref{singlefunctionmetric}. In fact, to prove it, one only needs to realise that junction condition \eqref{2.2 f(R)} implies that $\beta=\beta_0=\const$ Accordingly, equation \eqref{2.1 f(R)} becomes
\begin{equation} \label{ansantz2 eq}
    A_{t*}(2\beta_0^2-A_*)=0.
\end{equation}
The first option, $A_{t*}=0$, is discarded as per Result \ref{theostatic}, while the second one, $A_*=2\beta_0^2=\const$, leads to the inconsistent result $\dot{r}_*^2=-\beta_0^2<0$ when one resorts to equation \eqref{rp} for $\dot{r}_*$. As a result, we conclude that an exterior metric of the form \eqref{singlefunctionmetric} is not compatible with gravitational collapse in $f(R)$ theories of gravity with $f_{RR}(R)\neq 0$. Thus, Result \ref{theosinglefunction} is proved: no `single-function spacetime', either static or not, matches the dust star interior smoothly in non-linear $f(R)$ gravity. $\blacksquare$

As already mentioned in Section \ref{intandext}, Result \ref{theosinglefunction} conveys a generalisation of the theorem in \cite{Bueno:2017sui} stating that no \textit{static} `single-function spacetime' can be a exterior spacetime non-linear metric $f(R)$ gravity. Nonetheless, Result \ref{theosinglefunction} only holds provided that the interior spacetime is a dust-star FLRW metric, while the theorem in \cite{Bueno:2017sui} applies regardless of the interior matter source.

At this stage, a simple non-static ansatz satisfying the necessary condition $B(t,r)\neq 1/A(t,r)$ would be of the form \eqref{eqnResult4}, i.e.~a generalisation of `single-function' spacetimes in which we have included a $t$- and $r$-dependent redshift function $U(t,r)$. It is not difficult to show that this ansatz is also unsatisfactory: the combination of junction condition \eqref{2.1 f(R)} with the $\tau$ derivative of \eqref{2.2 f(R)} implies
\begin{equation}
    U_{r}=0\myskip\Rightarrow\myskip U(t,r)=U(t)\,.
\end{equation}
As a consequence, the metric \eqref{eqnResult4} becomes static, since $U(t)$ can always be absorbed in the differential of $t$ through a coordinate transformation $\sqrt{U(t)}\,\dif t\rightarrow\dif t$. Since static exteriors are ruled out by Result \ref{theostatic}, we have thus shown the validity of Result \ref{theoalmostANV}. $\blacksquare$

Another simple ansatz for time-dependent generalisations of the exterior metric would be as given in \eqref{eqnResult5}. However, this ansatz does not work either; equating \eqref{2.1 f(R)} with the $\tau$ derivative of \eqref{2.2 f(R)} one obtains
\begin{equation}
    \left(2\beta_0^2-\dfrac{U}{V}\right)V_{t}=0\,.
\end{equation}
This equation is satisfied if either $V(t)=\const$ or if $U(r)/V(t)=2\beta_0^2\Rightarrow U(r)=\const,\,V(t)=\const$; in both cases, the metric becomes static. As a result, by virtue of Result \ref{theostatic}, line elements of the form \eqref{eqnResult5} do not satisfy the junction conditions of $f(R)$ gravity. This is precisely the content of Result \ref{theoanothernonstatic}. $\blacksquare$

In order to conclude, let us mention  that  throughout this section we have been capable of imposing restrictive constraints on the exterior spacetime. In particular, our results indicate that, in non-linear metric $f(R)$ gravity, the spacetime outside a collapsing uniform-density dust star, if it exists, must be of the form \eqref{sphericallysymmetric}, with highly non-trivial ---and probably non-separable--- functions $A(t,r)$ and $B(t,r)\neq 1/A(t,r)$. Thus, the exterior in these theories seems to be substantially different from the sole Schwarzschild metric appearing in GR, yet the former should somehow reduce to the latter in the appropriate limits. Given the fact that most renowned solutions of non-linear metric $f(R)$ gravity are either constant-curvature spacetimes or static, Results \ref{theoconstcurv} and \ref{theostatic} rule out such exterior solutions as viable for matching FLRW-like, spatially-uniform dust-star interiors. For example, one of the most promising candidates, the Clifton I solution (also dubbed the Clifton-Barrow spacetime) \cite{Clifton:2005aj,Clifton:2006ug} fails to comply with the junction conditions because it is static.

\section{A new static solution of metric $f(R)$ gravity} 
\label{staticsol}

Abandoning the assumption that the star is collapsing, it is possible to find a previously undiscovered solution of a large class of metric $f(R)$ gravity models using the junction conditions studied above. More concretely, this new spacetime is a solution of every $f(R)$ theory satisfying $f(0)=0$ and $f_R(0)=0$.

Our starting point is the matching between the dust star FLRW interior \eqref{FLRW} and a static exterior spacetime of the form \eqref{MetricStaticAB} which will be a solution for some ---still unspecified--- $f(R)$ gravity model. As seen in Section \ref{Proofs Results 1 2 Corollary}, junction condition \eqref{static beta dot} is satisfied provided that
\begin{equation}
\label{BNewSolution}
    B(r)=\dfrac{1}{\beta_0^2}=\const,
\end{equation}
although this constraint together with \eqref{1.1 f(R)} and \eqref{rp} would imply that the interior solution is also static:
\begin{equation}
    \dot{a}=\dfrac{\dot{r}_*}{\chi_*}=0.
\end{equation}
Using the standard normalisation $a(0)=1$ (which entails that $r_*=\chi_*=\const$), coordinate $\chi$ reduces to the areal radius $r$. The interior metric then becomes
\begin{equation} \label{staticint}
    \dif s_-^2=\dif\tau^2-\dfrac{\dif r^2}{1-k r^2}-r^2\dif\Omega_{(2)}^2
\end{equation}
and, as a straightforward consequence, both $f(R^-)$ and the stellar density $\rho=\rho_0$ become constant in $\tau$ as well. The trace of the $f(R)$ equations of motion for the interior spacetime \eqref{staticint} reads
\begin{equation} 
\label{static trace}
    f_R^-R^- -2f^-=-\kappa T^-=-\kappa\rho_0\,.
\end{equation}
Assuming that $R^-$ is a variable and not a value, expression \eqref{static trace} can be interpreted as a differential equation for $f(R^-)$. This equation may be immediately integrated, revealing that \eqref{staticint} is a solution of
\begin{equation} \label{static theory}
    f(R)=\alpha R^2+\dfrac{\kappa\rho_0}{2},
\end{equation}
where $\alpha$ is an integration constant.\footnote{Notice that the statement that the correct theory is \eqref{static theory} is stronger than the assertion that equation \eqref{static trace} holds for \eqref{staticint}, since \eqref{static theory} would accomplish \eqref{static trace} for every (interior) solution, not just \eqref{staticint}. Moreover, as we shall see later, the spurious dependence of $f(R)$ on $\rho_0$ disappears when one takes into account the (vacuum) equations of motion for the exterior solution, which force $\rho_0=0$.}

On the other hand, should the interior solution have constant curvature scalar $R^-$, the full set of equations of motion of $f(R)$ gravity \eqref{f(R) EOM} become
\begin{equation} \label{EOM static theory}
    f_R^-R_{\mu\nu}^- -\dfrac{f^-}{2}g_{\mu\nu}^-=-\kappa T_{\mu\nu}^-.
\end{equation}
Evaluating equations \eqref{EOM static theory} for theory \eqref{static theory} together with the static, constant-curvature interior metric \eqref{staticint}, one finds that equations \eqref{EOM static theory} only impose one additional constraint relating $k$ and $\rho_0$ as follows:
\begin{equation}
    k^2=\dfrac{\kappa\rho_0}{24\alpha}.
\end{equation}

We shall now see how the exterior solution is fixed by the remaining junction conditions ---the third and the fourth ones, i.e.~equations \eqref{3 f(R)} and \eqref{all t and r 4 f(R)}, respectively---. Since $B(r)$ is given by \eqref{BNewSolution}, then
equation \eqref{all t and r 4 f(R)} is automatically satisfied, while equation \eqref{3 f(R)} becomes
\begin{equation}
    R^+=\dfrac{6kr_*^2}{r^2}=\dfrac{6(1-\beta_0^2)}{r^2}.
\end{equation}
The left-hand side of this expression is obtained by substituting \eqref{BNewSolution} in the expression for $R^+$:
\begin{equation}
    R^+=\dfrac{2(1-\beta_0^2)}{r^2}-\beta_0^2\left(\dfrac{A_{rr}}{A}-\dfrac{A_r^2}{2A^2}+\dfrac{2}{r}\dfrac{A_r}{A}\right).
\end{equation}
Equating the two previous expressions, and solving the resulting ordinary differential equation for $A(r)$, one finds
\begin{equation}
    A(r)=D\,r^{-(\Delta+1)}\,(C+r^{\Delta})^2,
\end{equation}
where
\begin{equation}
    \Delta\equiv\sqrt{9-\dfrac{8}{\beta_0^2}},
\end{equation}
while $C$ and $D$ are integration constants. Notice that $D$ can be absorbed inside $\dif t^2$ by a redefinition of the time coordinate $Dt\rightarrow t$. In what follows, we may thus set, without loss of generality, $D=1$. Bearing all of this in mind, the exterior metric becomes
\begin{equation} \label{staticext}
    \dif s^2_+=r^{-(\Delta+1)}\,(C+r^{\Delta})^2\,\dif t^2 -\dfrac{\dif r^2}{\beta_0^2} -r^2\,\dif \Omega_{(2)}^2.
\end{equation}
By construction, metrics \eqref{staticint} and \eqref{staticext} satisfy all the junction conditions at $r=r_*=\const$ Moreover, we have found that \eqref{staticint} solves the field equations of $f(R)$ gravity provided that $f(R)=\alpha R^2+\kappa\rho_0/2$. Consequently, we must confirm whether the exterior spacetime \eqref{staticext} is a vacuum solution for this class of $f(R)$ models. By inserting the exterior line element \eqref{staticext} in the vacuum equations of motion of \eqref{static theory}, one finds that the latter are only satisfied provided that $\beta_0^2=1$, which, in turn, implies
\begin{equation} \label{Delta = 1 consequences}
    \Delta=1,\myskip k=0\,\Rightarrow\,\rho_0=0.
\end{equation}
Therefore, we see that the matching is only possible if the $f(R)$ models \eqref{static theory} reduce to
\begin{equation} \label{R2}
    f(R)=\alpha R^2,
\end{equation}
i.e.~purely quadratic $f(R)$ gravity.

Because of \eqref{Delta = 1 consequences}, the interior solution  reduces to Minkowski spacetime. The exterior metric \eqref{staticext}, however, becomes
\begin{equation} \label{novelstatic}
    \dif s^2_+=\left(1+\dfrac{C}{r}\right)^2\,\dif t^2 -\dif r^2-r^2\,\dif \Omega_{(2)}^2.
\end{equation}
Therefore, the exterior metric, which is reminiscent of the static Clifton I solution \cite{Clifton:2006ug}, remarkably remains non-trivial for $C\neq 0$.\footnote{Let us note that $C$ is a parameter which remains free even after the matching is done.} Additionally, \eqref{novelstatic} can be shown to be Ricci flat, i.e.~$R^+=0$,\footnote{The fact that $R^+=0$ for this solution explains why this metric matches Minkowski spacetime at any $r_*$, cf.~\eqref{3 f(R)} and \eqref{4 f(R)}.} even though it has a non-vanishing Ricci tensor, i.e.~$R_{\mu\nu}^+\neq 0$. Thus, line element \eqref{novelstatic} constitutes an example of constant-curvature solution of $f(R)$ gravity which is not a solution of GR.

Due to its vanishing scalar curvature, our novel metric \eqref{novelstatic} turns out to be not only a solution of $f(R)=\alpha R^2$, but of \textit{any} theory satisfying $f(0)=0$ and $f_R(0)=0$ \cite{Nzioki:2009av,Calza:2018ohl}, such as the so-called `power of GR' models \cite{Clifton:2005aj} $f(R)\propto R^{1+\delta}$ having $\delta>0$.\footnote{The Clifton I or Clifton-Barrow spacetime is also a solution of these `power of GR' theories.} Therefore, the following highly non-evident result can be established:

\begin{result} \label{ResultVacuole}
    The vacuole consisting of an interior Minkowski spacetime smoothly matched to \eqref{novelstatic} at any areal radius $r=r_*=\const$ is a properly matched solution of any $f(R)$ model satisfying $f(0)=0$ and $f_R(0)=0$, i.e.~it satisfies both the equations of motion and all of the junction conditions of those $f(R)$ theories.
\end{result}

To the best of our knowledge, this vacuole solution is one of the first known examples of properly glued solutions of $f(R)$ gravity which is not a solution of GR (recall that $R_{\mu\nu}\neq 0$ for the outer solution). In addition, solution \eqref{novelstatic} has several interesting properties. For instance, direct computation of the Kretschmann scalar yields
\begin{equation}
    \mathcal{K}=R_{\mu\nu\rho\sigma}R^{\mu\nu\rho\sigma}=\dfrac{24 C^2}{r^6}\left(1+\dfrac{C}{r}\right)^{-2},
\end{equation}
revealing that \eqref{novelstatic} ---without the Minkowski interior--- has a curvature singularity $r=0$. Furthermore, if $C<0$, this spacetime has a curvature singularity at $r=-C$. The singularity at $r=0$ is always cured by matching \eqref{novelstatic} to a Minkowski interior at any $r_*>0$. Similarly, if $C<0$, the singularity at $r=-C$ can be cured if the vacuole extends at least up to $r=-C$. Other interesting features of this spacetime will be explored in future works.

\section{Conclusions and future prospects}
\label{Conclusions}

Aware of the fact that the junction conditions in $f(R)$ gravity are much more restrictive than their Einsteinian counterparts, the present investigation has paved the way for finding the appropriate generalisation of Oppenheimer-Snyder collapse to $f(R)$ gravity. Such a generalisation is  not only interesting in itself; via transformation to the Einstein frame, it would shed light on the issue of the de-scalarization of matter on scalar-tensor theories, i.e.~on the precise mechanism by which matter could get rid of its scalar hair so as to form a hairless black hole, as required by the no-hair theorems.

In the bulk of the text, we have developed the general formalism allowing one to determine whether a spherically symmetric exterior solution matches a dust star FLRW-like interior in $f(R)$ gravity. There are indeed two ways of tackling the problem: either one employs these junction conditions \eqref{1.1 f(R)}--\eqref{4 f(R)}, \eqref{tp} and \eqref{rp} directly in order to infer information about the exterior metric, or one simply inserts a known vacuum solution of $f(R)$ gravity into the junction conditions and checks whether such conditions are satisfied. Both procedures are not exempt from difficulties: analytic computations can be hard or even impossible to perform, while the way in which the problem could be tackled numerically remains unclear. Furthermore, there are very few known exact vacuum solutions of $f(R)$ gravity \cite{Faraoni:2021nhi}, a fact hampering the research on the topic.

Notwithstanding these shortcomings, the foundations of both approaches have been presented herein. We have also ruled out several classes of exterior solutions for the uniform-density dust star using the aforementioned junction conditions in Section \ref{ruling out}. In particular, we have proved that no constant-curvature exterior spacetime (either static or dynamic) can be smoothly matched to the dust star interior in non-linear $f(R)$ gravity (Result \ref{theoconstcurv}). Furthermore, we have also offered a rigorous proof of the result that no static exterior spacetime can be glued to the FLRW interior in non-linear $f(R)$ gravity (Result \ref{theostatic}).

The power of our formalism has allowed us to extend the known result that static spacetimes satisfying $g_{tt}g_{rr}=-1$ cannot be the exterior of any matter source in $f(R)$ gravity. Herein, we have shown that, at least in the case of a dust star interior, even a non-static exterior satisfying $g_{tt}g_{rr}=-1$ cannot match the interior FLRW-like spacetime (Result \ref{theosinglefunction}). We have also been able to establish further constraints on the exterior metric (Results \ref{theoalmostANV}, \ref{theoanothernonstatic} and the Corollary to Result \ref{theoconstcurv}).

Finally, to the best of our knowledge, we have found for the first time in the literature a new static vacuum solution of every $f(R)$ theory satisfying $f(0)=0$ and $f_R(0)=0$. The solution consists of a vacuole made out of an interior Minkowski spacetime surrounded by the previously undiscovered exterior \eqref{novelstatic}. The vacuole satisfies the four junction conditions of non-linear metric $f(R)$ gravity at any areal radius $r=r_*=\const$, and has a series of interesting properties, as discussed in Section \ref{staticsol}. For example, the curvature singularities of the exterior spacetime can be cured by the glueing to the Minkowski interior. We intend to provide a more detailed study on the characteristics of the novel solution in future works.

Is gravitational collapse of a uniform-density dust star possible in $f(R)$ theories of gravity? Even though this simple and illustrative model of gravitational collapse is not feasible in the Palatini version of the theory (c.f.~Appendix \ref{Appendix Palatini}), there are reasons to expect that it is still possible in the metric formulation of $f(R)$ gravity. However, we have shown that the mathematical description of the process must be highly non-trivial, convoluted, and radically different from the Oppenheimer-Snyder picture.

There is still plenty of work to be done so as to fully understand gravitational collapse in metric $f(R)$ gravity. Since the catalogue of possible suitable exteriors is meagre, looking for novel time-dependent solutions ---of the forms not ruled out by our results--- would certainly contribute to a better understanding of gravitational collapse in non-linear metric $f(R)$ gravity, as well as to a better understanding of the no-hair theorems for black-holes.

It is possible that the no-hair theorems could allow one to rule out exteriors by simple inspection, at least in principle. If the no-hair theorems hold, the exterior spacetime must \textit{dynamically} become Schwarzschild at some point. Therefore, spacetimes which do not reduce to Schwarzschild after some time\footnote{Depending on the coordinate system, this time could be infinite. For example, it is well-known that, in Oppenheimer-Snyder collapse, the Schwarzschild black hole takes infinite time to form as seen from the exterior; see, for instance, \cite{Weinberg:1972kfs}.} cannot describe gravity outside a collapsing non-rotating uniform-density dust star in $f(R)$ theories satisfying the no-hair theorems.

Furthermore, in $f(R)$ theories which do not satisfy the no-hair theorems, such as the so-called `power of GR' models $f(R) \propto R^{1+\delta}$, one does not know \textit{a priori} whether a given solution describes the gravitational field outside the star, the outcome of collapse, neither, or both. Coincidentally, the only solution which remains compatible with all of our results is ---as far as we are aware--- the Clifton II spacetime \cite{Clifton:2006ug}, which is a solution of these `power of GR' theories. The Clifton II solution is highly non-trivial \cite{Faraoni:2009xb}, and the analysis of the junction conditions following the lines of Appendices \ref{ArealRadiusJCAppendix} and \ref{NOTArealRadiusJCAppendix} seems to be impossible to perform analytically. For these reasons, further (numerical) studies are required in order to determine whether the Clifton II spacetime can be the exterior spacetime corresponding to a collapsing dust star. These future works will likely require a change in the way in which the problem is dealt with; for example, novel numerical techniques might be required.

\section{Acknowledgements}

We would like to thank Jose Beltr\'{a}n Jim\'{e}nez and Rituparno Goswami for their very insightful comments. This research is partly supported by the Spanish Ministerio de Ciencia e Innovaci\'{o}n (MICINN) under the research Grant No.~PID2019–108655 GB-I00. A.~C.-T. is supported by a Universidad Complutense de Madrid-Banco Santander predoctoral contract CT63/19-CT64/19. \'{A}.~d.~l.~C.-D.~acknowledges support from South African NRF Grants No.~120390, reference: BSFP190416431035; No.~120396, reference: CSRP190405427545; and Spanish MICINN Grants PID2019–108655 GB-I00 and COOPB204064, I-COOP+2019.

\appendix

\section{Oppenheimer-Snyder collapse} 
\label{Appendix A - Oppenheimer-Snyder}

Although unrealistic and purely academic in nature, the Oppenheimer-Snyder construction is the simplest possible description of gravitational collapse one can possibly devise. In their seminal 1939 paper \cite{Oppenheimer:1939ue}, Oppenheimer and Snyder considered a spatially uniform sphere of dust (i.e.~a presureless perfect fluid) collapsing under its own gravitational pull within the framework of GR. The absence of pressure inside the star implies that no other interaction aside from gravity is present. Therefore, the model is able to capture the essential features of gravitational collapse while retaining computational simplicity.

Because the matter that makes up the star is only subject to gravity, any fluid element falls freely, that is to say, following time-like geodesics. This allows one to introduce a coordinate system $x_-^\mu=(\tau,\chi,\theta,\varphi)$ adapted to such motion on the interior spacetime. Time coordinate $\tau$ represents the proper time along the geodesics, while $\chi$ is a comoving radial coordinate, i.e.~each fluid element is associated to a single fixed value of $\chi$ which remains unchanged during the entire process of collapse. This implies that the stellar surface $\Sigma_*$ is always located at
\begin{equation} \label{intstellarsurface}
    \chi=\chi_*=\const
\end{equation}
in these coordinates.

The line elements within and outside the star are determined by solving the Einstein equations with the corresponding matter sources (dust for the interior region and vacuum for the exterior). On the one hand, the metric inside the star turns out to be \cite{Weinberg:1972kfs} a closed ($k>0$) Friedmann-Lema\^{i}tre-Robertson-Walker (FLRW) spacetime,
\begin{equation} \label{FLRW}
    \dif s_-^2=\dif\tau^2-a^2(\tau)\left(\dfrac{\dif\chi^2}{1-k\chi^2}+\chi^2\dif\Omega_{(2)}^2\right),
\end{equation}
whose scale factor $a(\tau)$ satisfies the cycloid equation
\begin{equation} \label{cycloid}
    \dot{a}^2=k\left(\dfrac{1}{a}-1\right),
\end{equation}
where $\dot{\,}\equiv\dif/\dif\tau$, with initial condition $a(0)=1$ ---notice that this implies that $\dot{a}(0)=0$---. Constant $k>0$ is related to the initial energy density $\rho_0$ of the star,
\begin{equation} \label{k OS}
    k=\dfrac{\kappa\rho_0}{3},
\end{equation}
while the conservation equation requires such an energy density to evolve in $\tau$ as $\rho(\tau)=\rho_0/a^3(\tau)$.

On the other hand, in GR, the only possible exterior solution is, by virtue of Birkhoff's theorem \cite{Birkhoff}, the Schwarzschild metric,
\begin{equation} \label{Schwarzschild}
    \dif s_+^2=\left(1-\dfrac{2GM}{r}\right)\dif t^2-\left(1-\dfrac{2GM}{r}\right)^{-1}\dif r^2-r^2\dif\Omega_{(2)}^2,
\end{equation}
Coordinates $x_+^\mu=(t,r,\theta,\varphi)$ ---hereafter, the `exterior' coordinates--- are not comoving with the matter within the star. As a result, $\Sigma_*$ becomes $\tau$-dependent when expressed in exterior coordinates:
\begin{equation} \label{stellarsurface out r}
    t=t_*(\tau),\myskip r=r_*(\tau).
\end{equation}

The spacetime resulting from the matching of the interior and exterior metrics at $\Sigma_*$ will be a properly glued solution of the Einstein field equations provided that the interior and exterior spacetimes satisfy the Israel-Darmois junction conditions of GR \cite{Darmois,Israel:1966rt}, namely
\begin{itemize}
    \item \textbf{First junction condition}: the continuity of the induced metric $h_{ab}$ at $\Sigma_*$, i.e.~$[h_{ab}]=0$, and
    \item \textbf{Second junction condition}: the continuity of the extrinsic curvature $K_{ab}$ of $\Sigma_*$, i.e.~$[K_{ab}]=0$.
\end{itemize}
These two quantities are defined at either side of $\Sigma_*$ as
\begin{gather}
    h_{ab}^\pm=\dfrac{\partial x_\pm^\mu}{\partial y^a}\dfrac{\partial x_\pm^\nu}{\partial y^b}g_{\mu\nu}^\pm, \\
    K_{ab}^\pm=-n_\mu^\pm\left(\dfrac{\partial^2 x_\pm^\mu}{\partial y^a \partial y^b}+\Gamma^\mu_{\pm\alpha\beta}\dfrac{\partial x_\pm^\alpha}{\partial y^a}\dfrac{\partial x_\pm^\beta}{\partial y^b}\right), \label{Kab}
\end{gather}
where $y^a$ are the induced coordinates at $\Sigma_*$ and $n_\mu$ is the normal to that surface. In our case, the most convenient choice is $y^a=(\tau,\theta,\varphi)$.

For illustrative purposes, and to introduce notation, let us briefly review in the following how the junction conditions of GR give shape to the Oppenheimer-Snyder model of gravitational collapse, without detailing the computations that lead to the matching equations. For further reference (and completeness), the exhaustive derivation ---which closely follows the textbook treatment of the problem given in reference \cite{Poisson:2009pwt}--- may be found in Appendix \ref{ArealRadiusJCAppendix}, more precisely in its first three subsections. 

As in any other matching, if the junction conditions are satisfied, they must provide us with two crucial pieces of information:
\begin{itemize}
    \item how the matching surface $\Sigma_*$ evolves, in this case the evolution of $t_*$ and $r_*$ in proper time $\tau$, and
    \item whether there is a relationship between the parameters of the interior and exterior spacetimes, whic in the case of Oppenheimer-Snyder collapse are $k\propto\rho_0$ and $M$, respectively.\footnote{We must stress that $\chi_*$, the star's comoving radius, should not be regarded as a \textit{parameter} of the interior metric but an \textit{initial condition} for the matching: as per equation \eqref{1.1 f(R)}, $\chi_*=r_*(0)$ and is thus arbitrary ---for more information, see footnote \ref{chistar arbitrary}---. Nonetheless, $\chi_*$ will still appear in the equation relating $k\propto\rho_0$ and $M$, as one would intuitively expect.}
\end{itemize}
Hence, in order to understand not only how, but also why, the dust star interior matches the Schwarzschild exterior, we first need to decipher which of these two roles each junction condition plays in the glueing of both spacetimes.

As shown in Appendix \ref{Oppenheimer-Snyder Appendix}, the first junction condition ---$[h_{ab}]=0$--- will yield only two independent equations, while the second junction condition ---$[K_{ab}]=0$--- will only yield one. Let us first state the two constraints coming from the first junction condition, that is to say, the matching the induced metrics at $\Sigma_*$. These are \eqref{1.1 f(R)} and \eqref{1.2 f(R)}. Equation \eqref{1.1 f(R)} simply states that $r_*$ is proportional to the scale factor of the interior metric,\footnote{We note that we do not write $a_*$ in the junction conditions because $a$ is a function of $\tau$ only, and thus $a_*=a$.} which evolves in $\tau$ according to equation \eqref{cycloid} ---let us also remark that, upon differentiation, \eqref{1.1 f(R)} also provides the value of all the $\tau$ derivatives of $r_*$, in particular $\dot{r}_*$---. Because $a(0)=1$, this means that the stellar radius decreases from its initial value $r_*(0)=\chi_*$ ---i.e.~the star's comoving radius--- to zero in finite time, following cycloid curve \eqref{cycloid}.

On the other hand, equation \eqref{1.2 f(R)} becomes
\begin{equation}
    \dot{t}_*=\dfrac{\sqrt{\dot{r}_*^2+A_*}}{A_*}, \label{1.2 OS int}
\end{equation}
where $A(r)=1-2GM/r$, in the case of Schwarzschild. Because we already know the $\tau$-dependence $r_*$ and $\dot{r}_*$, expression \eqref{1.2 OS int} simply turns into an ordinary first order differential equation for $t_*$, which always has a solution given some initial condition. Consequently, the interpretation of the equations coming from the first junction condition is clear: they allow for a complete determination of the evolution in $\tau$ of the stellar surface as seen from outside the star, which is given by functions $r_*(\tau)$ and $t_*(\tau)$.

Having considered the first junction condition, we must also impose the second one, namely, the continuity of the induced metric at $\Sigma_*$. As previously mentioned ---and shown in Appendix \ref{Oppenheimer-Snyder Appendix}---, the matching of the extrinsic curvatures at the stellar surface only provides an additional equation, which is
\begin{equation} \label{rdot OS}
    \dot{r}_*^2=1-k\chi_*^2-A_*.
\end{equation}
$r_*$ and $\dot{r}_*$ may now be replaced in favour of $a$ and $\dot{a}$ using the first junction condition \eqref{1.1 f(R)}, yielding
\begin{equation}
    \dot{a}^2=-k+\dfrac{2GM}{a \chi_*^3}.
\end{equation}
Meanwhile, $\dot{a}$ may be expressed in terms of $a$ using the cycloid equation \eqref{cycloid}, leading to
\begin{equation}
    k\left(\dfrac{1}{a}-1\right)=-k+\dfrac{2GM}{a \chi_*^3}.
\end{equation}
We immediately notice that the $a$s cancel, and with them all the dependence in $\tau$ disappears from the previous expression. As a result, we finally obtain a relation between $M$, $k\propto\rho_0$ and $\chi_*$:
\begin{equation} \label{M OS}
    M=\dfrac{k}{2G}\chi_*^3=\dfrac{4\pi}{3}\rho_0\chi_*^3,
\end{equation}
where in the last step we have made use of \eqref{k OS}. This completes the Oppenheimer-Snyder construction in GR.

With minimal modifications, the Oppenheimer-Snyder model of gravitational collapse is also a properly matched solution of GR plus a cosmological constant $\Lambda$ \cite{Markovic:1999di}, which may be understood as $f(R)=R-2\Lambda$. This theory is special among all $f(R)$ models because, in this case, $f_{RR}(R)=0$ for any $R$, which implies that its junction conditions are exactly the same as in GR without a cosmological constant \cite{Deruelle:2007pt,Senovilla:2013vra}.

If a cosmological constant is included in the gravitational action, the exterior metric is necessarily (Anti-)de Sitter-Schwarzschild spacetime. As a result, equations \eqref{1.2 OS int} and \eqref{rdot OS} are also valid in this case, but now with $A(r)=1-2GM/r-\Lambda r^2/3$. The equation for the scale factor, however, gets modified if a cosmological constant is present, and becomes
\begin{equation}
    \dot{a}^2=\dfrac{\kappa\rho_0}{3a}+\dfrac{\Lambda a^2}{3}-k.
\end{equation}
Thus, the value of $k$ also changes in this theory:
\begin{equation}
    k=\dfrac{\kappa\rho_0+\Lambda}{3}.
\end{equation}
Combining these expressions with junction conditions \eqref{1.2 OS int} and \eqref{rdot OS}, one finds that the matching is possible for any $\Lambda$ provided that, once again, $M=4\pi\rho_0\chi_*^3/3$. However, in the presence of a cosmological constant, the star may either collapse or bounce depending on the specific values of $\Lambda$ and $M$. Moreover, if $M>\sqrt{\Lambda}/3$, the exterior spacetime cannot avoid contracting into a `big-crunch' singularity as well, dragged by the gravitational pull of the collapsing dust star. More complicated choices for function $f$ should produce similar effects. In particular, a modification of the scale factor dynamics and of the expression for $k$ is always to be expected in any $f(R)$ theory. As we have seen, the modified dynamics could potentially lead to a complete evasion of gravitational collapse, or even to the formation of singularities. Therefore, these possible issues must always be carefully considered.

\section{Junction conditions between the uniform-density dust star interior and a spherically symmetric exterior, using the areal radius as a coordinate} \label{ArealRadiusJCAppendix}

In this section, we shall obtain the junction conditions resulting from smoothly matching an interior FLRW spacetime \eqref{FLRW} with the most general spherically symmetric line element, across the time-like boundary $\Sigma_*$ given by \eqref{intstellarsurface} in interior coordinates $x_-^\mu=(\tau,\chi,\theta,\varphi)$.

For this calculation, we note that one can always choose `areal-radius' coordinates $x_+^\mu=(t,r,\theta,\varphi)$ such that the exterior metric takes the form
\begin{equation} \label{sphericallysymmetric appendix}
    \dif s^2_+=A(t,r)\,\dif t^2 -B(t,r)\,\dif r^2-r^2\,\dif \Omega_{(2)}^2,
\end{equation}
where $A$ and $B$ are two functions which completely characterise the exterior spacetime. In these `areal-radius' coordinates, the matching surface is given by expressions \eqref{stellarsurface out r}. We shall closely follow the treatment of the problem given in reference \cite{Poisson:2009pwt}.

\subsection{First junction condition}

On the one hand, as seen from the interior of the star,
\begin{equation} \label{dx-/dy}
    \dfrac{\partial x_-^\mu}{\partial y^a}=\delta^\mu_a,
\end{equation}
and thus the induced metric on the inner side of $\Sigma_*$ is
\begin{equation} \label{h-}
    \dif s_{\Sigma^-_*}^2=\dif\tau^2-a^2\chi_*^2\,\dif\Omega^2.
\end{equation}
On the other hand, from the exterior,
\begin{equation} \label{dx+/dy}
    \dfrac{\partial x_+^\mu}{\partial y^a}=(\dot{t}_*\,\delta^\mu_t+\dot{r}_*\,\delta^\mu_r)\,\delta^\tau_a+\delta^\mu_\theta\,\delta^\theta_a+\delta^\mu_\varphi\,\delta^\varphi_a.
\end{equation}
Consequently, the induced metric on the outer side of $\Sigma_*$ is given by
\begin{equation} \label{h+}
    \dif s_{\Sigma^+_*}^2=(A_*\dot{t}_*^2-B_*\dot{r}_*^2)\,\dif\tau^2-r_*^2\,\dif\Omega_{(2)}^2,
\end{equation}
where $A_*\equiv A(t_*(\tau),r_*(\tau))$ and $B_*\equiv B(t_*(\tau),r_*(\tau))$. The equality of the induced metrics \eqref{h-} and \eqref{h+} at both sides of $\Sigma_*$ imposes two conditions on the metric functions, namely \eqref{1.1 f(R)} and \eqref{1.2 f(R)}. Equation \eqref{1.2 f(R)} can be conveniently rearranged to produce expression \eqref{betadeff(R)}, which serves as the definition of function $\beta=\beta(\tau)$.

\subsection{Second junction condition} \label{JC 2 r Appendix}

In order to compute the extrinsic curvature at the junction surface, we first need to determine the unit normal to $\Sigma_*$, $n_\mu$. If $u^\mu$ denotes the four-velocity any fluid element, then $n_\mu$ is completely characterised by spherical symmetry together with the normalisation and orthogonality conditions $g^{\mu\nu} n_\mu n_\nu=0$ and $u^\mu n_\mu=0$ (respectively). In interior (comoving) coordinates,
\begin{equation}
    u_-^\mu=\dfrac{\partial x_-^\mu}{\partial\tau}=\delta^\mu_\tau,
\end{equation}
and $n_\mu^-$ is thus fixed to be
\begin{equation} \label{n-}
    n^-_\mu=\dfrac{a}{\sqrt{1-k\chi_*^2}}\delta^\chi_\mu.
\end{equation}

Combining \eqref{Kab}, \eqref{dx-/dy} and \eqref{n-} one finds that the extrinsic curvature of $\Sigma_*$, as seen from the inside, is
\begin{equation}
    K^-_{ab}=-\dfrac{a}{\sqrt{1-k\chi_*^2}}\Gamma^\chi_{-*ab}.
\end{equation}
Because the only relevant and non-vanishing Christoffel symbol of the interior FLRW spacetime is
\begin{equation}
    \Gamma^\chi_{-\theta\theta}=\dfrac{\Gamma^\chi_{-\varphi\varphi}}{\sin^2\theta}=-\chi\,(1-k\chi^2),
\end{equation}
the only non-zero components of $K^-_{ab}$ are
\begin{equation} \label{K-angularOS}
    K^-_{\theta\theta}=\dfrac{K^-_{\varphi\varphi}}{\sin^2\theta}=a\chi_*\sqrt{1-k\chi_*^2}.
\end{equation}

In exterior coordinates, the four-velocity of the fluid is
\begin{equation}
    u_+^\mu=\dfrac{\partial x_+^\mu}{\partial\tau}=\dot{t}_*\delta^\mu_t+\dot{r}_*\delta^\mu_r,
\end{equation}
so the (properly normalised) normal vector is
\begin{equation} \label{unormal+}
    n^+_\mu=\sqrt{A_* B_*}\,(-\dot{r}_*\delta^\mu_t+\dot{t}_*\delta^\mu_r),
\end{equation}
where we have also made a consistent choice of the overall sign. Hence, as seen from outside, the extrinsic curvature of $\Sigma_*$ becomes
\begin{eqnarray} \label{K+tr}
    \dfrac{K^+_{ab}}{\sqrt{A_* B_*}}&=&(\ddot{t}_*\dot{r}_*-\dot{t}_*\ddot{r}_*)\delta^\tau_a\delta^\tau_b
    +(\dot{r}_*\Gamma^t_{+*\alpha\beta}-\dot{t}_*\Gamma^r_{+*\alpha\beta})
    \nonumber\\
    &&\times\,\dfrac{\partial x_+^\alpha}{\partial y^a}\dfrac{\partial x_+^\beta}{\partial y^b}.
\end{eqnarray}
In order to compute the components of $K_{ab}^+$, we make use of \eqref{dx+/dy} and take into account that the only relevant, non-vanishing Chistoffel symbols are
\begin{equation}
    \begin{gathered}
    \Gamma^t_{+*tt}=\dfrac{A_{t*}}{2A_*},\myskip\Gamma^r_{+*rr}=\dfrac{B_{r*}}{2B_*},\myskip\Gamma^r_{+*\theta\theta}=-\dfrac{r_*}{B_*}, \\
    \Gamma^t_{+*tr}=\Gamma^t_{+*rt}=\dfrac{A_{r*}}{2A_*}, \myskip\Gamma^t_{+*rr}=\dfrac{B_{t*}}{2A_*}, \\
    \Gamma^r_{+*tr}=\Gamma^r_{+*rt}=\dfrac{B_{t*}}{2B_*},\myskip\Gamma^r_{+*tt}=\dfrac{A_{r*}}{2B_*}.
\end{gathered}
\end{equation}
After some calculations, we obtain equations
\eqref{2.1 f(R)} and \eqref{2.2 f(R)} corresponding to the second junction condition, with function $\beta$ and parameter $\beta_0$ being respectively given by \eqref{betadeff(R)} and \eqref{beta0 f(R)}.

\subsection{Interlude: Oppenheimer-Snyder collapse in GR} \label{Oppenheimer-Snyder Appendix}

As explained before, GR has only two junction conditions: the first one, $[h_{ab}]=0$, and the second one, $[K_{ab}]=0$. Therefore, the relevant junction conditions in this case are equations \eqref{1.1 f(R)} and \eqref{1.2 f(R)}, \eqref{2.1 f(R)} and \eqref{2.2 f(R)}. Moreover, in GR, the exterior can only be Schwarzschild,
\begin{equation}
    A(t,r)=\dfrac{1}{B(t,r)}=1-\dfrac{2GM}{r},
\end{equation}
and thus function $\beta$, defined in \eqref{betadeff(R)}, reduces to
\begin{equation} \label{beta OS appendix 1}
    \beta=\sqrt{\dot{r}_*^2+A_*}.
\end{equation}
The Schwarzschild metric is independent of $t$. As a result, \eqref{2.1 f(R)} yields $\dot{\beta}=0$. This is compatible with the other constraint coming from the second junction condition, \eqref{2.2 f(R)}, which reduces to
\begin{equation} \label{beta OS appendix 2}
    \beta=\beta_0=\sqrt{1-k\chi_*^2}.
\end{equation}
Combining \eqref{beta OS appendix 1} with \eqref{beta OS appendix 2}, one finds that, in Oppenheimer-Snyder collapse,
\begin{equation} \label{JC 2 OS}
    \dot{r}_*^2=\beta_0^2-A_*=-k\chi_*^2+\dfrac{2GM}{r_*}.
\end{equation}
Substituting \eqref{1.1 f(R)} and the equation for $a(\tau)$, which in GR is \eqref{cycloid}, one finally finds that junction conditions require
\begin{equation}
    M=\dfrac{k}{2G}\chi_*^3=\dfrac{4\pi}{3}\rho_0\chi_*^3.
\end{equation}

\subsection{Third and fourth junction conditions}

The first new junction condition arising from $f(R)$ gravity is the continuity of the Ricci scalar at $\Sigma_*$, i.e.~$[R]=0$.\footnote{It is worth mentioning that the Ricci scalar of the exterior solution is given in terms of functions $A$ and $B$ by
\begin{eqnarray*}
    R^{+}=&-&\dfrac{A_{rr}}{AB}+\dfrac{A_r}{2AB}\left(\dfrac{A_r}{A}+\dfrac{B_r}{B}\right)-\dfrac{2}{r}\left(\dfrac{A_r}{AB}-\dfrac{B_r}{B^2}\right) \nonumber\\
    &&+\,\dfrac{2}{r^2}\left(1-\dfrac{1}{B}\right)+\dfrac{B_{tt}}{AB}-\dfrac{B_t}{2AB}\left(\dfrac{A_t}{A}+\dfrac{B_t}{B}\right).
\end{eqnarray*}
} In terms of the scale factor of the interior FLRW spacetime, this condition reads
\begin{equation} \label{3 f(R) appendix partial}
    R^+_*=R^-_*=6\left(\dfrac{\dot{a}^2+k}{a^2}+\dfrac{\ddot{a}}{a}\right),
\end{equation}
Equation \eqref{1.1 f(R)} can be employed to reexpress \eqref{3 f(R) appendix partial} in terms of $r_*$ and its derivatives, yielding \eqref{3 f(R)}.

The other novel junction condition coming from $f(R)$ gravity is the continuity of the normal derivative of the Ricci scalar at the stellar surface, i.e.~$[n^\mu\partial_\mu R]=0$. As seen from inside the star, this normal derivative is simply
\begin{equation}
    g_{-*}^{\mu\nu}n_\mu^-\partial_\nu R^-_*=R^-_{\chi*}=0.
\end{equation}
As seen from the exterior, the normal derivative of $R_+$ does not vanish in principle:
\begin{equation}
    g_{+*}^{\mu\nu}n_\mu^+\partial_\nu R^+_*=-\sqrt{A_* B_*}\left(\dfrac{\dot{r}_*}{A_*}R_{r*}^++\dfrac{\dot{t}_*}{B_*}R_{t*}^+\right).
\end{equation}
This forces us to require \eqref{4 f(R)} for the fourth junction condition to be accomplished.

\section{Junction conditions between the uniform-density dust star interior and a spherically symmetric exterior, without using the areal radius as a coordinate}
\label{NOTArealRadiusJCAppendix}

The most general spherically symmetric exterior line element can always be expressed as
\begin{equation} \label{exteriorx}
    \dif s^2_+=C(\eta,\xi)\,\dif \eta^2 -D(\eta,\xi)\,\dif\xi^2-r^2(\eta,\xi)\,\dif \Omega_{(2)}^2.
\end{equation}
As we can see, the difference between \eqref{exteriorx} and \eqref{sphericallysymmetric appendix} is that the areal radius $r$ is not a coordinate, but a function of the new time variable $\eta$ and the new spatial coordinate $\xi$ instead. We intend to develop the junction conditions resulting from glueing \eqref{FLRW} and \eqref{exteriorx} across the stellar surface $\Sigma_*$ using coordinates $(\eta,\xi,\theta,\varphi)$, and then compare the results with those of Appendix \ref{ArealRadiusJCAppendix}.

As our starting point, we must note that the stellar surface is now given by
\begin{equation} \label{sigmax}
    \eta=\eta_*(\tau),\myskip \xi=\xi_*(\tau)
\end{equation}
in these coordinates. Additionally, we now have that
\begin{equation} \label{r(t,r)}
    r=r(\eta,\xi)\myskip\Rightarrow\myskip r_*(\tau)=r_*(\eta_*(\tau),\xi_*(\tau)).
\end{equation}

\subsection{First junction condition}

Since, as seen from outside the star,
\begin{equation} \label{dx+/dy x}
    \dfrac{\partial x_+^\mu}{\partial y^a}=(\dot{\eta}_*\,\delta^\mu_\eta+\dot{\xi}_*\,\delta^\mu_\xi)\,\delta^\tau_a+\delta^\mu_\theta\,\delta^\theta_a+\delta^\mu_\varphi\,\delta^\varphi_a,
\end{equation}
the induced metric on the exterior side of $\Sigma_*$ will be
\begin{equation} \label{h+ x}
    \dif s_{\Sigma^+_*}^2=(C_*\dot{\eta}_*^2-D_*\dot{\xi}_*^2)\,\dif\tau^2-r_*^2\,\dif\Omega_{(2)}^2,
\end{equation}
where $C_*\equiv C(\eta_*(\tau),\xi_*(\tau))$ and $D_*\equiv D(\eta_*(\tau),\xi_*(\tau))$. Equalling the induced metrics \eqref{h-} and \eqref{h+ x}, one obtains two conditions on the metric functions, namely equation \eqref{1.1 f(R)} ---which remains unchanged--- and
\begin{equation} \label{1.2 f(R) x}
    C_*\dot{\eta}_*^2-D_*\dot{\xi}_*^2=1.
\end{equation}

Equation \eqref{1.2 f(R) x} can be conveniently rearranged as
\begin{equation} \label{betadeff(R) x}
    C_*\,\dot{\eta}_*=\sqrt{C_*+C_*D_*\dot{\xi}_*^2}\equiv\tilde{\beta},
\end{equation}
which serves as the definition of function $\tilde{\beta}=\tilde{\beta}(\tau)$.

\subsection{Second junction condition}

The four-velocity of the fluid is now
\begin{equation}
    u_+^\mu=\dfrac{\partial x_+^\mu}{\partial\tau}=\dot{\eta}_*\delta^\mu_\eta+\dot{\xi}_*\delta^\mu_\xi,
\end{equation}
so the (properly normalised) normal vector is
\begin{equation} \label{unormal+ x}
    n^+_\mu=\sqrt{C_* D_*}\,(-\dot{\xi}_*\delta^\mu_\eta+\dot{\eta}_*\delta^\mu_\xi).
\end{equation}
Therefore, the extrinsic curvature of $\Sigma_*$ expressed in exterior coordinates is
\begin{eqnarray} \label{K+tr x}
    \dfrac{K^+_{ab}}{\sqrt{C_* D_*}}&=&(\ddot{\eta}_*\dot{\xi}_*-\dot{\eta}_*\ddot{\xi}_*)\delta^\tau_a\delta^\tau_b
    +(\dot{\xi}_*\Gamma^\eta_{+*\alpha\beta}-\dot{\eta}_*\Gamma^\xi_{+*\alpha\beta})
    \nonumber\\
    &&\times\,\dfrac{\partial x_+^\alpha}{\partial y^a}\dfrac{\partial x_+^\beta}{\partial y^b}.
\end{eqnarray}
Proceeding exactly as we did back in Appendix \ref{JC 2 r Appendix}, we obtain, after a rather long computation, the following two independent equations for the second junction condition:
\begin{gather}
    \dot{\tilde{\beta}}=\dfrac{C_{\eta*}\dot{\eta}_*^2-D_{\eta*}\dot{\xi}_*^2}{2}, \label{2.1 f(R) x} \\
    \tilde{\beta}=\dfrac{\beta_0\sqrt{C_* D_*}-r_{\eta*}D_*\dot{\xi}_*}{r_{\xi*}}, \label{2.2 f(R) x}
\end{gather}
where $\beta_0$ is given again by expression \eqref{beta0 f(R)}, $C_\eta\equiv\partial C/\partial\eta$, $D_\eta\equiv\partial D/\partial\eta$, $r_\eta\equiv\partial r/\partial\eta$ and $r_\xi\equiv\partial r/\partial\xi$.

\subsection{Third and fourth junction conditions}

It is almost immediate to check that the third junction condition still leads to equation \eqref{3 f(R)}.\footnote{Obviously, \eqref{r(t,r)} must be taken into account in this case. \eqref{r(t,r)} entails that $\dot{r}_*=r_{\eta*}\dot{\eta}_*+r_{\xi*}\dot{\xi}_*$, so the third junction condition is much more convoluted in these coordinates.} Finally, the fourth junction condition now yields\footnote{We would like to remark that, in these coordinates, the Ricci scalar of the exterior solution is given by
\begin{eqnarray*}
    R^{+}=&-&\dfrac{C_{\xi\xi}}{CD}+\dfrac{C_\xi}{2CD}\left(\dfrac{C_\xi}{C}+\dfrac{D_\xi}{D}\right)-\dfrac{2r_\xi}{r}\left(\dfrac{C_\xi}{CD}-\dfrac{D_\xi}{D^2}\right) \nonumber\\
    &&+\,\dfrac{2}{r^2}\left(1+\dfrac{r_\eta^2}{C}-\dfrac{r_\xi^2}{D}\right)+\dfrac{D_{\eta\eta}}{CD}-\dfrac{D_\eta}{2CD}\left(\dfrac{C_\eta}{C}+\dfrac{D_\eta}{D}\right)\nonumber\\ &&+\,\dfrac{4}{r}\left(\dfrac{r_{\eta\eta}}{C}-\dfrac{r_{\xi\xi}}{D}\right)-\dfrac{2}{r^2}\left(\dfrac{C_\eta}{C^2}-\dfrac{D_\eta}{CD}\right).
\end{eqnarray*}}
\begin{equation} \label{4 f(R) x}
    \dfrac{\dot{\xi}_*}{C_*}R_{\xi*}^++\dfrac{\dot{\eta}_*}{D_*}R_{\eta*}^+.
\end{equation}
As we can clearly see, the junction conditions for \eqref{exteriorx} reduce to \eqref{1.1 f(R)}--\eqref{2.2 f(R)}, \eqref{3 f(R)} and \eqref{4 f(R)} if one sets $r=\xi$ and performs the substitutions $\eta\rightarrow t$, $C\rightarrow A$ and $D\rightarrow B$, since $\tilde{\beta}$ also reduces to $\beta$ ---i.e.~to expression \eqref{betadeff(R)}--- under these circumstances.

\subsection{Summary and comparison with the results of Appendix \ref{ArealRadiusJCAppendix}}

To sum up, the relevant junction conditions arising from the smooth matching of a FLRW dust star interior \eqref{FLRW} and \eqref{exteriorx} at \eqref{sigmax} are equations \eqref{1.1 f(R)}, \eqref{3 f(R)}, \eqref{1.2 f(R) x} and \eqref{2.1 f(R) x}--\eqref{4 f(R) x}. These constraints are to be complemented with the definition of $\tilde{\beta}$, equation \eqref{betadeff(R) x}.

Junction conditions \eqref{1.2 f(R) x} and \eqref{2.1 f(R) x}--\eqref{4 f(R) x} are more complex than their counterparts \eqref{1.2 f(R)}--\eqref{2.2 f(R)} and \eqref{4 f(R)}. This was to be expected, since coordinates $(\eta,\xi,\theta,\varphi)$ are more general than $(t,r,\theta,\varphi)$. As a result, computations should be ---in principle--- much more difficult to perform when the exterior spacetime is expressed as in \eqref{exteriorx}.

For example, the expressions for $\dot{\eta}_*$ and $\dot{\xi}_*$ in terms of $C_*$ and $D_*$ are not as simple as the expressions \eqref{tp} and \eqref{rp} for $\dot{t}_*$ and $\dot{r}_*$ in terms of $A_*$ and $B_*$. This is because equation \eqref{2.2 f(R) x} includes a term proportional to $\dot{\xi}_*$ in its right-hand side ---in contrast, \eqref{2.2 f(R)} does not contain any term proportional to $\dot{r}_*$---. We thus have a quadratic equation for $\dot{\xi}_*$ after substituting \eqref{2.2 f(R) x} in \eqref{betadeff(R) x}. From this quadratic equation one obtains the following two solutions for $\dot{\xi}_*$:
\begin{equation} \label{xistar}
    \dot{\xi}_*=\dfrac{\sqrt{C_*D_*}\left(r_{\eta*} \beta_0\pm r_{\xi*}\sqrt{\dfrac{r_{\eta*}^2+C_*\beta_0^2}{D_*}-\dfrac{C_*}{D_*^2}r_{\xi*}^2}\right)}{D_* r_{\eta *}^2-C_* r_{\xi *}^2}.
\end{equation}
Substituting \eqref{xistar} in \eqref{betadeff(R) x} yields the two corresponding solutions for $\dot{\eta}_*$. Again, it is straightforward to check that both solutions for $\dot{\xi}_*$ and $\dot{\eta}_*$ respectively reduce to \eqref{rp} and \eqref{tp} if one sets $r=\xi$ and performs the substitutions $\eta\rightarrow t$, $C\rightarrow A$ and $D\rightarrow B$. Nonetheless, the highly convoluted appearance of $\dot{\xi}_*$ and $\dot{\eta}_*$ implies that using them to build a system of equations for metric functions $C$, $D$ and $r$ is in principle much more complicated than obtaining a system of equations for functions $A$ and $B$ using \eqref{rp} and \eqref{tp}, as we did back in Section \ref{ruling out}.

Accordingly, we clearly see that `areal-radius' coordinates $(t,r,\theta,\varphi)$ are more natural, in the sense that $r_*$ is always the quantity which becomes proportional to the interior scale factor $a$ as per the first junction condition \eqref{1.1 f(R)} ---which remains unmodified when one abandons `areal-radius' coordinates and switches to $(\eta,\xi,\theta,\varphi)$---. Junction conditions are also more difficult to handle when one uses the alternative coordinate system $(\eta,\xi,\theta,\varphi)$. This could potentially cause problems when the exterior spacetime cannot be analytically cast in the form \eqref{sphericallysymmetric appendix} using a coordinate transformation.

\section{Incompatibility of Oppenheimer-Snyder collapse with Palatini $f(R)$ gravity}
\label{Appendix Palatini}

The junction conditions of Palatini $f(R)$ gravity (in which $R$ is now the Ricci scalar of the independent connection) are different from those of GR and of metric $f(R)$ gravity. More precisely, according to \cite{Olmo:2020fri}, the relevant constraints in Palatini $f(R)$ gravity are
\begin{gather}
    [h_{ab}]=0\,, \label{Palatini JC 1} \\
    [T]=0\,, \label{Palatini JC 2} \\
    [K_{ab}]-\dfrac{1}{3}h_{ab}[K]=\dfrac{\kappa}{f_R}\tau_{ab}\,, \label{Palatini JC 3} \\
    \tau=0\,, \label{Palatini JC 4}
\end{gather}
in the case allowing for thin shells to be present, where $\tau_{ab}$ is the thin shell's stress-energy tensor (i.e.~the divergent part of the stress-energy tensor), $\tau\equiv h^{ab}\tau_{ab}$ is its trace, and $K\equiv h^{ab}K_{ab}$ is the trace of the extrinsic curvature of the matching surface $\Sigma$.

The case in which there are no thin shells at the matching surface is recovered by setting $\tau_{ab}=0$. It is then immediate to see that \eqref{Palatini JC 4} holds automatically, while condition \eqref{Palatini JC 3} becomes
\begin{equation}
    K_{ab}=\dfrac{1}{3}h_{ab}[K],
\end{equation}
whose trace is satisfied automatically for all $K_{ab}$. Conditions \eqref{Palatini JC 1} and \eqref{Palatini JC 2} are unaffected by the choice $\tau_{ab}=0$.

Condition \eqref{Palatini JC 2} suffices to show that Oppenheimer-Snyder collapse is impossible in Palatini $f(R)$ gravity. A dust star has an energy-momentum tensor whose trace is $T^-=\rho\neq 0$, while the exterior vacuum solution has $T^+=0$. Therefore, the trace of the stress-energy tensor cannot be continuous at the matching surface unless $\rho=0$, and thus the glueing with any dust star interior is impossible as per the junction conditions of Palatini $f(R)$ gravity derived in  \cite{Olmo:2020fri}. What is more, if the interior stress-energy tensor is that of a perfect fluid, then the matching is only possible provided that $\rho_*=3p_*$, i.e.~that the equation of state evaluated at the stellar surface is that of radiation. This is of course true if the fluid \textit{is} radiation, but also if one requires the equation of state to be such that $\rho_*=0$ entails $p_*=0$. This is typically the case in more realistic stars, but not in those made of dust, in which $p=0$ everywhere and the star can end abruptly at any given radius. The generic incompatibility of dust star interiors with junction condition \eqref{Palatini JC 2}, shows that the Oppenheimer-Snyder collapse model is not viable within Palatini $f(R)$ gravity:

\begin{result}
    Isolated bodies made of pressureless matter are incompatible with the junction conditions of Palatini $f(R)$ gravity presented in \cite{Olmo:2020fri}. Therefore, the Oppenheimer-Snyder model of gravitational collapse is also incompatible with such junction conditions.
\end{result}

Nonetheless, we must stress that one can still study gravitational collapse in Palatini $f(R)$; however, a different equation of state for the star is required by the junction conditions \eqref{Palatini JC 1}--\eqref{Palatini JC 4}. For example, non-dust perfect fluids and polytropic stars are still allowed by \eqref{Palatini JC 2}. It seems reasonable to expect, though, that a change in the equation of state might significantly complicate the mathematical treatment of the problem.

\nocite{*}
\bibliography{bibliography}

\end{document}